\documentclass[preprint]{aastex}
\usepackage{psfig}

\begin{document}

\title{Correlated X-ray Spectral and Timing Behavior of the Black Hole
Candidate XTE J1550--564: A New Interpretation of Black Hole States}

\author{Jeroen Homan\altaffilmark{1},
        Rudy Wijnands\altaffilmark{1,2},
        Michiel van der Klis\altaffilmark{1},
        Tomaso Belloni\altaffilmark{1,3},
        Jan van Paradijs$^\dagger$\altaffilmark{1,4},
        Marc Klein-Wolt\altaffilmark{1},
        Rob Fender\altaffilmark{1}, and
        Mariano M\'endez\altaffilmark{1,5}
}

\altaffiltext{1}{Astronomical Institute 'Anton Pannekoek', University
of Amsterdam, and Center for High Energy Astrophysics, Kruislaan 403,
1098 SJ, Amsterdam, The Netherlands}

\altaffiltext{2}{MIT, Center for Space Research, Cambridge, MA 02139 -- Chandra Fellow}

\altaffiltext{3}{Osservatorio Astronomico di Brera, Via E. Bianchi 46,
I-23807, Merate (LC), Italy}

\altaffiltext{4}{University of Alabama in Huntsville, Department of
Physics, Hunstville, AL 35899}

\altaffiltext{5}{Facultad de Ciencias Astron\'omicas y
Geof\'{\i}sicas, Universidad Nacional de La Plata, Paseo del Bosquey
S/N, 1900 La Plata, Argentina}

\begin{abstract}
We present an analysis of data of the black hole candidate and X-ray
transient XTE J1550--564, taken with the {\it Rossi X-ray Timing
Explorer} between 1998 November 22 and 1999 May 20. During this period
the source went through several different states, which could be
divided into soft and hard states based on the relative strength of
the high energy spectral component.  These states showed up as
distinct branches in the color-color and hardness-intensity diagrams,
connecting to form a structure with a comb-like topology, the branch
corresponding to the soft state forming the spine and the branches
corresponding to the various hard states forming the teeth of the
comb.

The power spectral properties of the source were strongly correlated
with its position on the branches. The broad band noise became
stronger, and changed from power law like to band limited, as the
spectrum became harder.  Three types of quasi-periodic oscillations
(QPOs) were found: 1--18 Hz and 102--284 Hz QPOs on the hard branches,
and 16--18 Hz QPOs on and near the soft branch. The 1--18 Hz QPOs on
the hard branches could be divided in three sub-types. The frequencies
of the high and low frequency QPOs on the hard branches were
correlated with each other, and anti-correlated with spectral
hardness. The changes in QPO frequency suggest that the inner disc
radius only increases by a factor of 3--4 as the source changes from a
soft to a hard state.
 
Our results on XTE J1550--564 strongly favor a two-dimensional
description of black hole behavior, where the regions near the spine
of the comb in the color-color diagram can be identified with the high
state, and the teeth with transitions from the high state, via the
intermediate state (which includes the very high state) to the low
state, and back. The two physical parameters underlying this
two-dimensional behavior vary to a large extent independently and
could for example be the accretion rate through the disk and the size
of the Comptonizing region causing the hard tail. The difference
between the various teeth is then associated with the mass accretion
rate through the disk, suggesting that high state $\leftrightarrow$
low state transitions can occur at any disk mass accretion rate and
that these transitions are primarily caused by another, independent
parameter. We discuss how this picture could tie in with the
canonical, one-dimensional behavior of black hole candidates that has
usually been observed.
\end{abstract}

\keywords{accretion, accretion disks -- black hole physics -- stars:
individual (XTE J1550--564) -- X-rays: stars }

\section{Introduction}\label{intro_sec}
The X-ray transient XTE J1550--564 was discovered on 1998 September 7
\citep{sm1998} with the {\it All Sky Monitor} (ASM) on board the {\it
Rossi X-ray Timing Explorer} (RXTE). Soon after optical
\citep{orbaja1998} and radio \citep{camchu1998} counterparts were
identified. Observations with the RXTE {\it Proportional Counter
Array} (PCA) were performed on an almost daily basis between 1998
September 7 and 1999 May 20.

\begin{figure}[t]
\centerline{\psfig{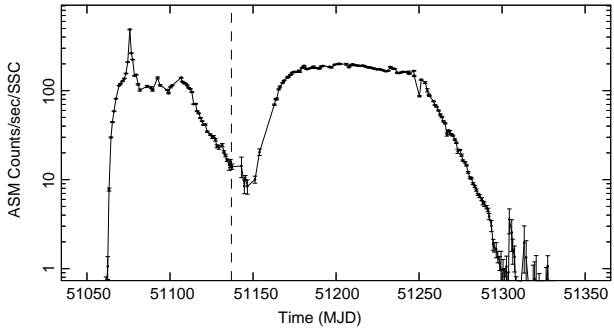}}
\caption{\scshape \small The one day averaged All Sky Monitor (ASM) 2--12 keV light
curve of XTE J1550--564. The dashed line marks the beginning of the
PCA data set used in this paper. \label{asm_fig}}
\end{figure}

The 1998/1999 outburst of XTE J1550--564 consisted of two parts,
separated by a minimum that occurred around 1998 December 3 (MJD
51150; see Figure \ref{asm_fig}).  The first part of the outburst
(until MJD 51150) has been the subject of work by \citet[timing
behavior during the rise]{cuzhch1999}, \citet[spectral
behavior]{somcre1999}, and \citet[timing behavior]{remcso1999}. During
the initial 10 day rise a quasi-periodic oscillation (QPO) was found,
together with a second harmonic \citep{cuzhch1999}.  It had a
frequency between 82 mHz and 4 Hz, that smoothly increased with the
X-ray flux. During the strong flare (reaching 6.8 Crab) that occurred
around MJD 51075 \citep{remomc1998}, a QPO was found with a frequency
of 185 Hz \citep{remcso1999}. High frequency oscillations were also
found on three other occasions, with frequencies between 161 and 238
Hz.  Low frequency QPOs (3--13 Hz) were also observed, during the
strong flare and the decay of the first part of the
outburst. Correlations between spectral parameters and the low
frequency QPOs (for the entire outburst) have been presented by
\citet{somcre2000a}.

Traditionally the behavior of black hole X-ray transients has been
described in terms of transitions between four canonical black hole
states (for overviews of black hole spectra and power spectra we refer
to \citet{tale1995} and \citet{va1995b}). The classification of an
observation into one of these four states is based on luminosity,
spectral and timing properties, and on the order in which they occur.

The spectra of black hole X-ray binaries are often described in terms
of a disk black body component, believed to be coming from an
accretion disk, and a power law tail at high energies, which is
thought to arise in a Comptonizing region (e.g. \citet{suti1980}). The
power spectra can be described by a (broken) power law, with sometimes
one or more quasi-periodic oscillation peaks superimposed. The
parameter usually thought to determine the state of the black hole is
the mass accretion rate. The definitions of the different states are
rather loose and have shown some variation between authors; we
therefore only give a general overview of the four canonical states in
order of inferred increasing mass accretion rate:

\begin{itemize}
\item{Low State (LS): The 2--20 keV X-ray spectrum can be described by
a single power law, with a photon index ($\Gamma$) of $\sim$1.5 plus
sometimes a weak disk black body component ($kT<$1 keV; less than a
few percent of the total luminosity). The power spectrum shows strong
band-limited noise with a typical strength of 20--50\% rms and a break
frequency ($\nu_{b}$) below 1 Hz.}
\item{Intermediate State (IS): In the X-ray spectrum both the power
law ($\Gamma\approx$2.5) and disk black body components ($kT\la$1 keV)
are present. The noise in the power spectrum is weaker (typically
5--20\% rms) and the break frequency is higher ($\nu_{b}\approx$1--10
Hz) than in the LS. QPOs between 1 and 10 Hz are sometimes observed.}
\item{High State (HS): The X-ray spectrum is dominated by the disk
black body component ($kT\approx$1 keV), and the power law component
($\Gamma\approx2-3$) is weak or absent. The noise in the power
spectrum is power law like and very weak, with a strength of less than
2--3 \% rms.}
\item{Very High State (VHS): Like in the IS the X-ray spectrum is a
combination of a disk black body ($kT\approx$1--2 keV) and a power law
($\Gamma\approx2.5$). The power spectrum shows noise, that can either
be HS-like (power law) or LS-like (band limited; 1-15\% rms,
$\nu_{b}\approx$1--10 Hz). QPOs are often seen in the VHS with
frequencies between 1 and 10 Hz.}
\end{itemize}

Note that there is little difference between the power spectral and
X-ray spectral properties of the VHS and IS, although conventionally
the total flux in the VHS is described as much higher than that in the
IS \citep{bemeva1996,meva1997,bevale1997}. The reason that the IS was
introduced as a separate state \citep{bevale1997} basically is that it
occurred in GS 1124--68 at epochs when the source appeared to be
evolving gradually from HS to LS and had a flux that was only 10\% of
that during the peak of the VHS \citep{ebogao1994}. It could not,
therefore, in the one-dimensional classification outlined above, be
identified with the VHS which by definition occurs at the upper end of
the inferred mass accretion range, above the HS. All three states, LS,
IS, and VHS are characterized by the presence of strong band limited
noise and a hard power law component, and are in that sense much more
similar to each other than to the HS, which is characterized by these
features being very weak or absent.

According to \citet{somcre1999} XTE J1550--564 went through the VHS,
HS, and IS during the first part of the outburst.  In this paper we
present a study of the correlated spectral and timing behavior of XTE
J1550--564 during the second part of its outburst.  We will discuss
the results for XTE J1550--564 using some of the canonical
terminology, in order to compare our results with those of other
transients. However, we will also discuss the discrepancies of the
canonical one-dimensional model with the results obtained for XTE
J1550--564; these discrepancies concern in particular the way in which
the various states relate to each other. We find the source moved
through all the four black hole states, in a way that is highly
suggestive of a new two dimensional interpretation of the black hole
states. 

In Section \ref{obs_sec} we explain our analysis methods. Our results
are presented in Sections \ref{lc-cd_sec}, \ref{power_sec}, and
\ref{radio_sec}. These sections are intended for the specialized
reader, the level of detail is rather high. A separate summary of the
most important results is therefore given at the beginning of the
discussion (Section \ref{discuss_sec}). We end by summarizing our main
conclusions in Section \ref{conlude_sec}.

\section{Observations and analysis}\label{obs_sec}
For our analysis we used all Public Target Of Opportunity RXTE/PCA
\citep{brrosw1993,jaswgi1996} data for XTE J1550--564 taken between
1998 November 22 23:38 UTC (MJD 51139)and 1999 May 20 19:37 UTC (MJD
51318). This adds up to 171 single observations, corresponding to a
total observing time of $\sim$400 ks. When we refer to a single
observation, we mean a part of the data with its own unique RXTE
observation ID; all observations will be referred to by their Modified
Julian Date (MJD) at the start of the observation. In all light curves
and color-color diagrams each point represents one single observation.

\begin{table}[t]
\begin{tabular}{llll}
Date & Time Resolution (s) &  \# Energy Channels & Energy Band (keV) \\
\hline
\hline
22/11/1998--20/05/1999 & $2^{-3}$ (Standard 1) & 1 & 2--60 \\
(MJD 51139--51318)     & $2^4$   $\,$ (Standard 2) & 128 & 2--60 \\
\hline
22/11/1998--22/03/1999 & $2^{-8}$  & 8 & 2--13.1 \\
(MJD 51139--51259)     & $2^{-13}$ & 1 & 2--6.5  \\
                       & $2^{-13}$ & 1 & 6.5--13.1 \\
                       & $2^{-16}$ & 1 & 13.1--60 \\
\hline                       
22/03/1999--29/04/1999 & $2^{-8}$  & 8 & 2--15.8 \\
(MJD 51259--51297)     & $2^{-13}$ & 1 & 2--7.9  \\
                       & $2^{-16}$ & 1 & 7.9--15.8 \\
                       & $2^{-16}$ & 1 & 15.8--60 \\
\hline
30/04/1999--20/05/1999 & $2^{-20}$ & 256 & 2--60 \\
(MJD 51297--51318)\\
\hline
\end{tabular}
\caption{\scshape \small Data modes of the RXTE Proportional Counter Array (PCA). The
first column gives the dates during which the different modes were
active (MJDs in brackets). The Standard 1 and 2 modes  (top two lines) were always
active.\label{modes_tab}}
\end{table}

The PCA data were obtained in several different modes (see Table
\ref{modes_tab}), some of which were active simultaneously.  On 1999
March 22 (MJD 51259), the high voltages of the PCA instrument were
changed, resulting in a different energy gain. The count rates and the
colors obtained after this change (gain Epoch 4) can not be directly
compared with the data obtained before the change (gain Epoch 3). Data
obtained during satellite slews, Earth occultations, and South
Atlantic Anomaly passages were removed from our data set.

The Standard 2 data (see Table 1) were used to create light curves,
color curves, color-color diagrams (CDs), and a hardness-intensity
diagram (HID). Only data of proportional counter units (PCUs) 0 and 2
were used for this, since these were the only two that were active
during all the observations. All PCA count rates and colors given in
this paper are only for those two PCUs combined. The data were
background subtracted, but dead time corrections ($<$6\%) were not
applied. For the light curves and colors, the photon energy channel
boundaries were chosen in such a way that the corresponding energies
for Epoch 3 and 4 matched as well as possible. A color is the ratio of
the count rates in two energy bands.  We define the soft color (SC) as
the ratio of the count rates in the 6.2--15.7 keV and 2--6.2 keV bands
(Epoch 3), or 6.1--15.8 keV and 2--6.1 keV bands (Epoch 4); hard color
(HC) is defined as the ratio of the count rates in the 15.7--19.4 keV
and 2--6.2 keV bands (Epoch 3), or 15.8--19.4 keV and 2--6.1 keV bands
(Epoch 4). This definition of colors, with the same band in the
denominator for both the hard and soft color, has the advantage that
comparison with a two-component model is straightforward.  Namely, if
the source spectrum is dominated by the contribution from two spectral
components (in our case a disk black body and a power law), then a
color data point will lie on the line connecting the color points of
the individual spectral components \citep{wa1982,vave1994} and the
ratio of distances from the data point to the points representing the
components is the inverse ratio of their contribution in the 2--6.2
keV (Epoch 3) or 2--6.1 keV (Epoch 4) bands. For the HID we used the
2--60 keV count rate (representing the full energy range covered by the PCA)
 as intensity, and the hard color (see above) as hardness.
The observation starting at MJD 51298.18 was included in the light
curves, but not in the CD, HID, and color curves, since at high
energies the source could not be detected above the background.

The Standard 2 data were also used to perform a number of spectral
fits. The spectra were background subtracted, and fitted in the
2.5--25.0 keV (Epoch 3) or 3.1--25.0 keV (Epoch 4) band, using a
systematic error of 2\%. Fits were made with XSPEC 10.00, using a fit
function that consisted of a disk black body, a power law, a Gaussian
line with a fixed energy of 6.5 keV and a width of 1--1.5 keV, and an
edge around 7 keV. Interstellar absorption was modelled using the
Wisconsin cross sections \citep{momc1983}, with $N_H$ fixed to a value
of 2$\times10^{22}$ atoms/cm$^2$ \citep{somcre1999}. We found that the
results of the spectral fits were very sensitive to the version of the
PCA response matrix we used. The response matrices were initially
created using FTOOLS version 4.2 and later with the updated version
5.0. Our initial fits showed that the inner disk radius and color
temperature of the disk black body component were, respectively,
correlated and anti-correlated with the hardness of the
spectrum. However, both correlations were found to be reversed when
using the updated version of the response matrices. In view of this we
decided to omit the spectral fits from the current paper, and only
discuss the spectral behavior using the color-color diagrams (which
are matrix-independent). For a complete spectral analysis of XTE
J1550--564 we refer to \citet{somcre1999,somcre2000c}.

The three high time resolution modes (with time resolution
$\le2^{-13}$ s, see Table \ref{modes_tab}) were used to produce
1/16--512 Hz power spectra in their respective energy bands and in the
combined 2--60 keV band; the same was done for the $2^{-20}$ s mode
(MJD 51297--51318), which was split into three energy bands with
similar energy ranges.  In order to study the variability at lower
frequencies, 1/128--128 Hz power spectra were created in the 2--60 keV
band, and for some observations also in 8 energy bands covering 2--60
keV. No background or dead time corrections were applied to the data
before the power spectra were created; the effect of dead time on the
Poisson noise was accounted for in the power spectral fits. The power
spectra were selected on time, count rate, color, or a combination of
these, before they were averaged and fitted. Although most of the
power spectra presented in this paper are normalized according to
\citet{ledael1983}, the actual power spectral fits were made to power
spectra that were rms normalized \citep{va1995a}.

The power spectra were fitted with a combination of several
functions. A constant was used to represent the Poisson level. The
noise at low frequencies was fitted with a power law
($P\propto\nu^{-\alpha}$), or with a broken power law
($P\propto\nu^{-\alpha_1}$ for $\nu<\nu_{b}$;
$P\propto\nu^{-\alpha_2}$ for $\nu>\nu_{b}$). In practice, the low
frequency noise component in the power spectra of most observations
could be fitted with a single power law. However, when combining
several observations, due to the smaller uncertainties, it became
apparent that a single power law did not yield acceptable fits,
especially around 1 Hz; using a broken power law for those combined
observations resulted in much better fits. For the single observations
we continued using a single power law, since for a broken power law
the break frequency was poorly constrained, and $\chi^2$ did not
differ significantly between the two fit functions. Most QPOs were
fitted with a Lorentzian, $P\propto [(\nu-\nu_c)^2 +
(FWHM/2)^2)]^{-1}$, where $\nu_c$ is the central frequency and FWHM
the full-width-at-half-maximum. In some cases narrow QPOs were found
for which a Lorentzian provided inadequate fits; in those cases a
Gaussian was used, $P\propto e^{(\nu-\nu_c)^2/\sigma^2}$, where
$\nu_c$ is the central frequency and $\sigma$ the width of the
Gaussian. Furthermore, we sometimes used an exponentially cutoff power
law ($P\propto\nu^{-\alpha} e^{-\nu / \nu_{cutoff}}$) to fit an extra
noise component at low frequencies. The errors on the fit parameters
were determined using $\Delta\chi^2=1$. The energy dependence of the
power spectral features was in general obtained by fixing all
parameters, except the amplitude, to their values obtained in a
specific band. However, in some cases, when the shape of the QPOs was
found to change between energy bands, the FWHM and/or frequency were
not fixed.  Upper limits on the strength of the power spectral
features were determined by fixing all their parameters, except the
amplitude, to values obtained in another energy band or observation,
and using $\Delta\chi^2=2.71$ (95\% confidence).

Unless otherwise stated, all the power spectral parameters are those
in the 2--60 keV band, and the noise rms amplitude is that in the
0.01--1 Hz range.

\begin{figure}
\centerline{\psfig{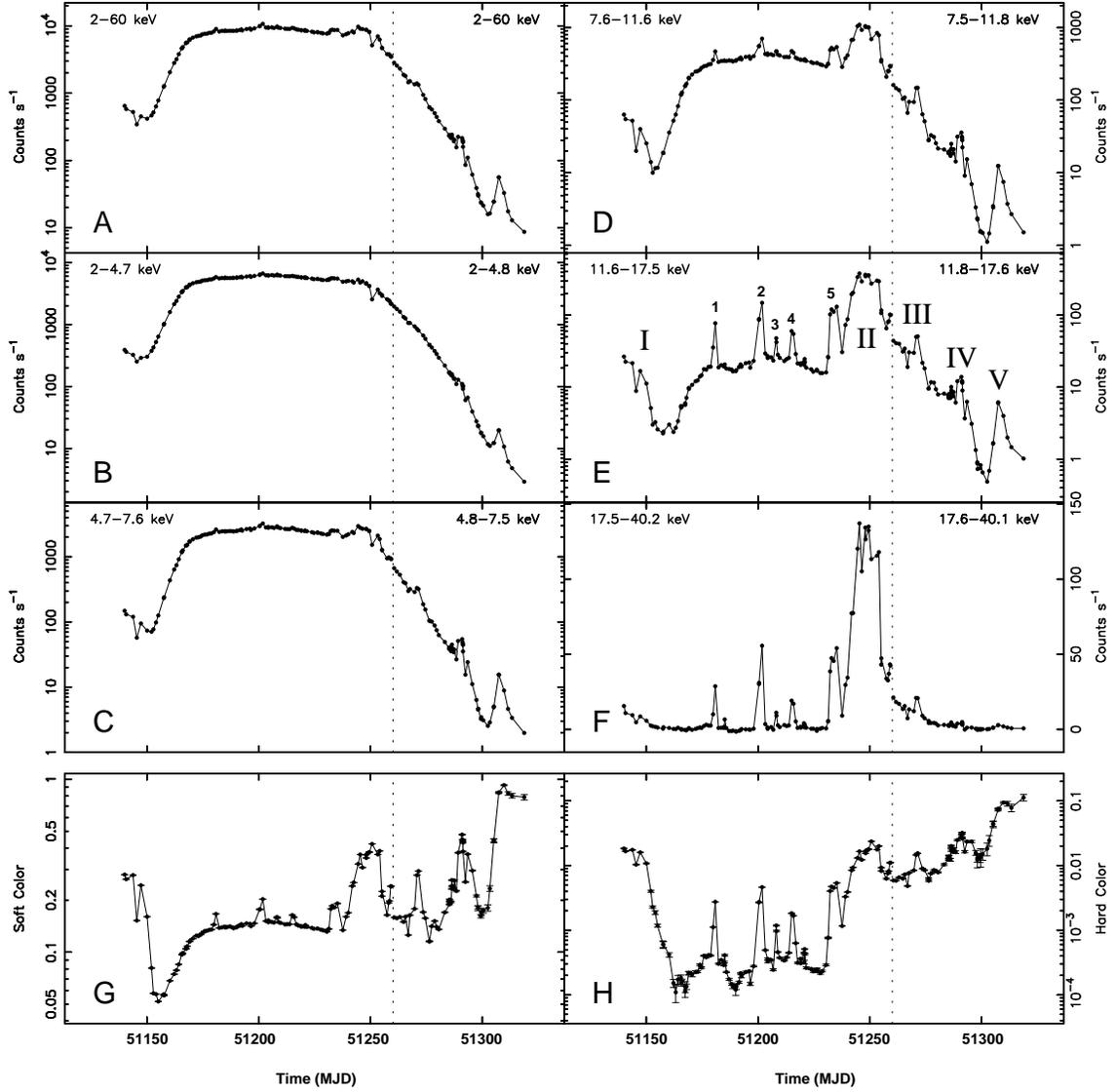}}
\caption{\scshape \small The background subtracted PCA light curves in six energy
bands (a--f), and the soft (g) and hard color (h) curves. The dotted
line indicates the date of the PCA gain change (see Section
\ref{obs_sec}). Energy bands are given for Epoch 3 (left) and Epoch 4
(right). Note that all curves, except (f), are plotted
logarithmically. In (e) the locations of the flares (1--5) and
branches (I--V) are indicated. Errors in the light curves are smaller
than the symbol size, and are therefore omitted. See Section \ref{obs_sec} for the
energy bands used to create the color curves.\label{lc_fig}}
\end{figure}

\section{Light curves, color--color diagrams}\label{lc-cd_sec}

Figure \ref{asm_fig} shows the one day averaged ASM 2--12 keV light
curve of the 1998/1999 outburst of XTE J1550--564, with the dashed
line marking the start of our PCA data set. It shows a broad local
minimum around MJD 51150, which naturally divides the outburst into
two parts. Following this minimum ($\sim$8.5 ASM counts s$^{-1}$) the
count rate increased by a factor of 10 within 20--25 days, and then
rose to about 200 counts s$^{-1}$ in 40 days. After a relatively flat
top XTE J1550--564 showed an initially slow decline ($\sim$55 days) to
about 100 ASM counts $s^{-1}$, which was followed by a decrease by a
factor of 100 in $\sim$45 days. On one day during during the first
part of the decline, MJD 51250, the ASM lightcurve showed a strong
dip; the count rate was found to be $\sim$40\% lower than in the two
adjacent observations (see also Section \ref{type-spec_sec}).  At the
end of the outburst, around MJD 51310, two small flares occurred,
reaching a few ASM counts $s^{-1}$. Although XTE J1550--564 never
again reached the level of the MJD 51074--51076 flare (6.8 Crab or
$\sim$490 ASM counts s$^{-1}$), it was bright (in the 2--12 keV band)
for a longer period of time during the second part than during the
first part of the outburst; during the first part it was observed
above 150 counts s$^{-1}$ on only six days, during the second part it
was above this level on more than 70 days.

The PCA data set used in our analysis started on MJD 51138 (dashed
line in Figure \ref{asm_fig}), just before the minimum between the two
parts of the outburst was reached. The PCA light curves in different
energy bands are shown in Figures \ref{lc_fig}a-f. The time of the PCA
gain change is indicated by the dotted line. As expected, the overall
2--60 keV light curve, dominated by the contribution of the low energy
bands, has a shape that is similar to that of the ASM light
curve. Note that as the photon energy increases, the local minimum
near MJD 51150 occurs later and seems to become broader. While the
light curves in the low energy bands have more or less the same
profile as in the ASM light curve, in the high energy bands they look
strikingly different. The light curve in Figure \ref{lc_fig}e shows
several strong flares on top of the overall outburst profile, and
above 17.5 keV (Fig. \ref{lc_fig}f) the light curve is dominated by
these flares. For later use the relatively small flares were numbered
1 to 5 (Fig. \ref{lc_fig}e), and the bigger (and broader) ones, that
clearly showed up as branches in the CD (see below) were numbered I to
V.

Figures \ref{lc_fig}g and \ref{lc_fig}h show the evolution of soft and
hard color with time. The change from Epoch 3 to 4 is again indicated
by a dotted line. Until the end of flare/branch II the colors were
quite well correlated with the count rate in all energy bands, the
only clear exception being the drop in hard color during the rise (MJD
51150--51160). After that, while the count rates dropped, the colors
increased (indicating a hardening of the spectrum) and were only
correlated with the count rates during flares/branches III-V (see
Fig. \ref{lc_fig}e).

\begin{figure}[t]
\centerline{\psfig{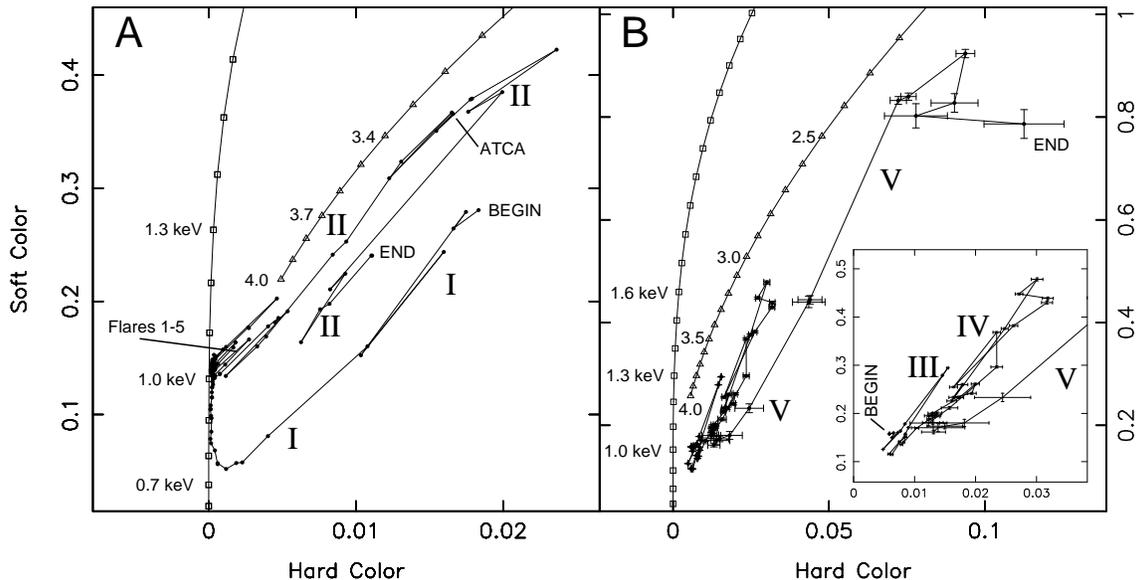}}
\caption{\scshape \small Color-color diagrams for RXTE Gain Epoch 3 (a) and Epoch 4
(b). The inset in (b) shows an enlargement of the lower left part of
the track in (b).  Errors in (a) are omitted since they were smaller
than the symbol size. The definition of the colors can be found in
Section \ref{obs_sec}.  The squares give the expected location for a
pure disk black body spectrum, with different temperatures as
indicated, and the triangles those for a pure power law spectrum, with
different indices as indicated. The begin and end points of the data
are marked. The locations of the flares and hard branches are
indicated by 1--5 and I--V, respectively.  The RXTE/PCA observation
taken on the day of our radio observation (MJD 51248) is indicated in
(a) by 'ATCA'.
\label{cd_fig}}
\end{figure}

Combining the two color curves, two color-color diagrams (CDs) were
produced. The CDs for Epoch 3 and 4 are shown in Figure \ref{cd_fig}a
and \ref{cd_fig}b, respectively.  In both CDs we also plotted the
expected colors for a disk black body (DBB) spectrum at different
temperatures (squares), and the expected colors for a power law
spectrum with different indices (triangles), both for an assumed $N_H$
of $2\times10^{22}$ atoms cm$^{-2}$ \citep{somcre1999}. Note that the
values of the expected colors (unlike the observed colors) do depend
on the version of the PCA response matrix that is used; the lines in
Figure \ref{cd_fig} were produced using FTOOLS version 4.2.  As
explained in Section \ref{obs_sec}, if the energy spectrum were a
combination of only a DBB and a power law, then the corresponding
point in the CD would lie on the straight line that connects the
appropriate point on the DBB curve with the one on the power law
curve. Although fits show that more spectral components are needed to
obtain a good $\chi^2$ (see e.g. \citet{somcre1999}), we do gain some
insight into how the parameters (temperature of the DBB, index of the
power law) and relative strength of the two most important spectral
components behave.

\begin{figure}[t]
\centerline{\psfig{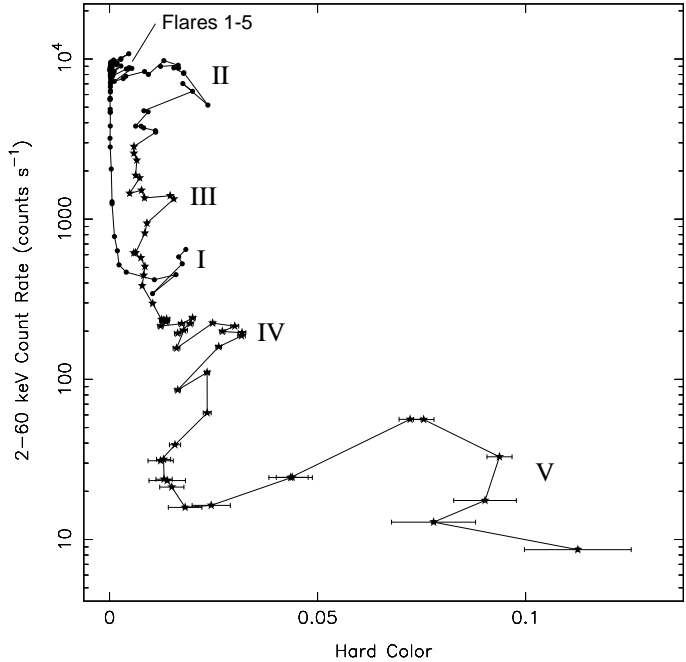}}
\caption{\scshape \small Hardness--Intensity diagram of the PCA data, with the 2--60
keV count rate as intensity. Bullets are RXTE Gain Epoch 3 data, stars Epoch 4
data. The location of the flares (1-5) and hard branches (I-V) is
given. See Section \ref{obs_sec} for definition of hard color.\label{hid_fig}}
\end{figure}

The pattern traced out in the CD before the gain change (Figure
\ref{cd_fig}a) shows roughly three branches. A spectrally soft branch
lies very close and nearly parallel to the DBB curve between 0.8 and
1.05 keV, and two spectrally hard branches (I and II, corresponding to
big flares I and II in Figure \ref{lc_fig}) that lie more or less
parallel to the power law curve. It should be noted that the values
for the temperature are read from the CD; values obtained from
spectral fits yield a maximum temperature of $\sim$1.1 keV (instead of
1.05 keV), see \citet{somcre2000c}. In the following, we give a short
description of how the source moved through the CD. The spectrum of
the first observation, on MJD 51139 (marked {\scriptsize BEGIN} in
Figure \ref{cd_fig}a), can be described as a combination of a DBB and
a power law spectrum; it neither lies close to the DBB curve nor to
the power law curve.  As time progressed the source moved to the left
in the CD along branch I, i.e. towards a pure DBB spectrum. Around MJD
51157 the source was located close to the DBB curve, with a
temperature of $\sim$0.8 keV. Subsequently the temperature increased
to $\sim$1.0 keV, while the source stayed close to the DBB
curve. Around MJD 51179, at SC$\sim$0.13 in the CD, the source
suddenly left the DBB curve in the direction of the power law curve
(flare 1), indicating a relative increase in the strength of the power
law component. On MJD 51182 the source had returned close to the DBB
curve, with a somewhat higher temperature than before. After that, the
temperature increased to a maximum value of $\sim$1.05 keV (on MJD
51204), and then decreased to $\sim$1.0 keV (on MJD 51231). During
this period four more flares (4--5) occurred, around MJDs 51200,
51208, 51215, and 51233 (see also Figure \ref{lc_fig}e). Similar to
flare 1, these four flares also pointed away from the DBB curve.
After each flare, except flare 5, the source returned to the DBB curve
at a similar temperature as before. During the decay of flare 5 a new
(big) flare started, which developed into branch/flare II. This branch
lay relatively close to the power law curve, indicating that the power
law became the dominant spectral component. The spectrally hardest
observation on branch II was the one on MJD 51250. After MJD 51250,
the source moved back into the direction of the DBB curve again, but
at a lower temperature. The time it took the source to move down
branch II was shorter than for it to move up that branch, $\sim$4.5
days and $\sim$9 days, respectively. The transition down from HC=0.02
to HC=0.008 even occurred in less than one day.  After the gain
change, on MJD 51260 (see inset in Figure \ref{cd_fig}b, indicated by
{\scriptsize BEGIN}) the source was found relatively close to the DBB
curve. A small branch (III) was traced out around MJD 51272, almost
parallel to the power law curve. From MJD 51283 to 51298 another
branch (IV) was traced out which also lay parallel to the power law
curve. Finally, during the last observations (MJDs 51299--51318) a
branch (V) was traced out, that pointed towards the regions of the
power law curve with indices lower than 2.5.

The HID (2--60 keV count rate vs. hard color) is shown in Figure
\ref{hid_fig}. It clearly shows that the five flares (1--5) occurred
at the highest count rates, and that the five branches (I--V) are well
separated from each other in count rate. On each hard branch the count
rate never varied more than by a factor 2--3. Note that at the lowest
count rates the observations between  hard branches tend to have
harder spectra than those at higher count rates.

\section{Power spectra}\label{power_sec}
The power spectra will be presented in order of time, but when
it seems more appropriate also according to their position in the
CD. We start by giving a short overview of the broad band power
spectral behavior in the next paragraph. Then, a more detailed study
follows of the power spectra during the start of the second part of
the outburst, the broad maximum and flares, branch II, and the decay.

\begin{figure}[t]
\centerline{\psfig{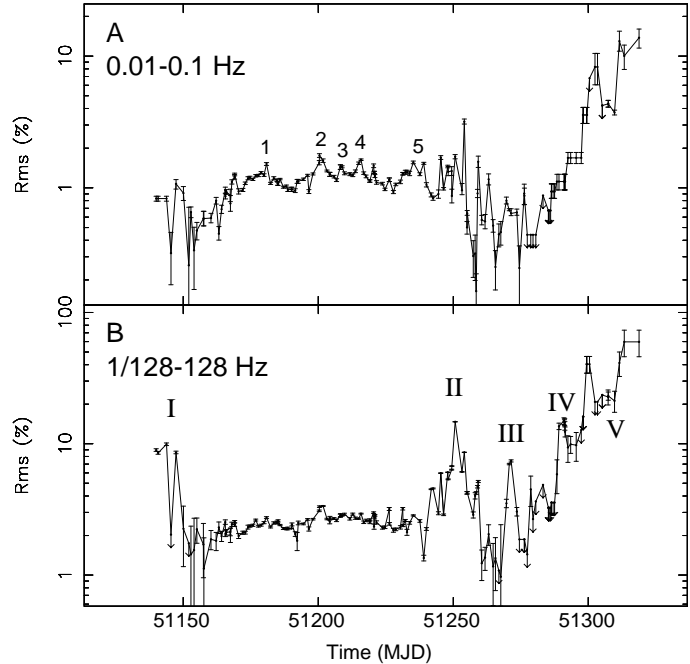}}
\caption{\scshape \small The fractional rms amplitude of 0.01--0.1 Hz noise (a) and
the 1/128--128 Hz noise (b) as a function of time. Some observations
were added together, to obtain significant detections, and they were
each plotted at their combined value. Upper limits are depicted by
small arrows. Flares are labelled 1--5 in (a) and branches I--V in
(b). \label{noise_curve_fig}}
\end{figure}

In Figure \ref{noise_curve_fig}b we show the total power in the
1/128--128 Hz power spectra as a function of time. One can immediately
see that increases in source variability occur whenever the source is
on one of the five branches. Between MJDs 51150 and 51240, when the
X-ray spectrum was soft, the total power had a strength of 1\% to 3\%
rms, which is typical for the high state. Similar weak noise was also
found between branches II, III, and IV. On branches I, II, and III the
power spectra had strengths between 5\% and 15\% rms, suggesting that
the source was is the intermediate and/or very high state. On the last
two branches the power increased from around 5\% to almost 60\%
rms. Noise rms amplitudes of several tens of percent are usually only
found in the low state.

\subsection{Start of the second part of the outburst}\label{rise}

\begin{figure}[t]
\centerline{\psfig{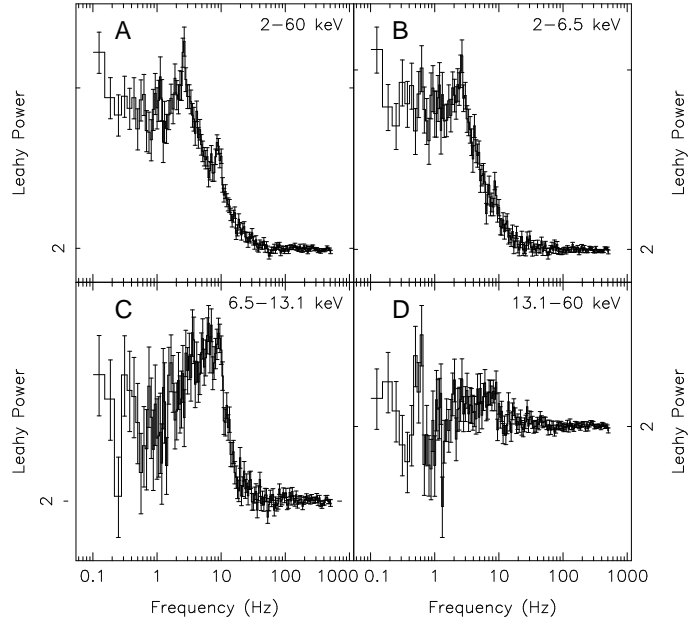}}
\caption{\scshape \small The combined power spectrum of the observations on MJD
51139--51143 in four energy bands. Poisson level was not
subtracted.\label{obs1-3_pow_fig}}
\end{figure}

\begin{figure}[t]
\centerline{\psfig{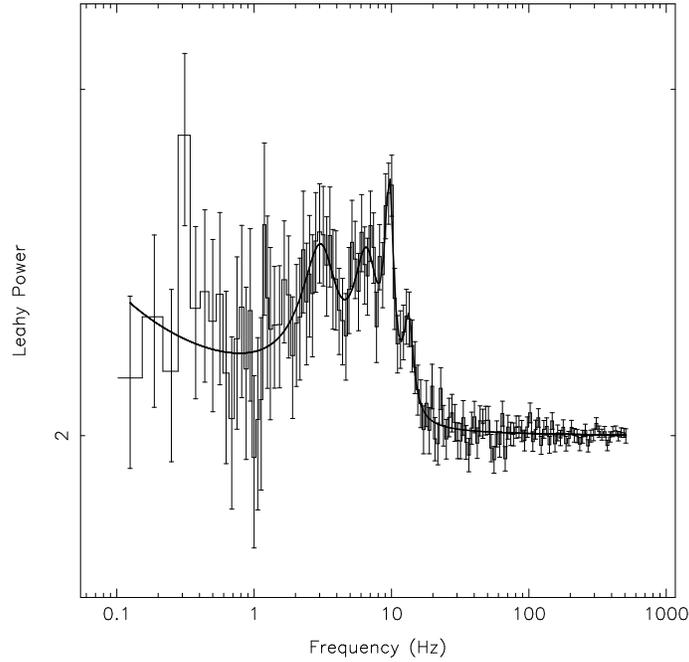}}
\caption{\scshape \small The 6.5--13.1 keV power spectrum of the MJD 51139. The solid
line shows the best fit with four QPOs and a power law. Poisson level
was not subtracted.\label{obs1_fit_fig}}
\end{figure}

During our first PCA observations the source was still in the decay of
the first part of the outburst. The power spectra of the first three
observations (MJDs 51139--51143) were very similar. Their combined
2--60 keV power spectrum (Figure \ref{obs1-3_pow_fig}a) showed band
limited noise at low frequencies, a peaked feature around 2.5 Hz and a
QPO around 9 Hz, which were fitted with, respectively, a broken power
law, and two Lorentzians.  The strength of the noise was
2.49$\pm$0.07\% rms, with $\nu_{b}=4.2^{+0.3}_{-0.6}$ Hz,
$\alpha_1=0.1^{+0.3}_{-0.6}$, and $\alpha_2=1.2\pm0.1$. The peaked
component or QPO close to the break had a frequency of 2.60$\pm$0.06
Hz, a FWHM of 1.0$\pm$0.5 Hz, and an rms amplitude of
2.4$\pm$0.6\%. The QPO at 9.17$\pm$0.14 Hz had a FWHM of 3.6$\pm$0.8
Hz, and an rms amplitude of 3.8$^{+0.4}_{-0.3}$\%. The power spectrum
depended strongly on energy, as can be seen from Figure
\ref{obs1-3_pow_fig}. At low energies (Fig. \ref{obs1-3_pow_fig}b) it
was dominated by the noise component with a peaked feature around the
break, whereas at higher energies (Fig. \ref{obs1-3_pow_fig}c) both
this peak and the noise component were replaced by a broad peak around
9 Hz.  The moderate noise strength and presence of QPOs classify these
observations as IS/VHS. Since the count rates in these observations
are considerably lower than during earlier (and later) observations
where QPOs and moderate noise strengths were found, the source was
probably in the IS rather than in the VHS. These observations were
also classified as IS by \citet{somcre1999}. It should be noted that
by combining several power spectra, narrow features in the individual
power spectra may be smoothed out and form broad bumps like the one
seen in Figure \ref{obs1-3_pow_fig}c. For instance, Figure
\ref{obs1_fit_fig} shows the 6.5--13.1 keV power spectrum of the first
of these three observations, on MJD 51139. It could be fitted with
QPOs at 3$\pm$0.2, 6.5$\pm$0.3, 9.63$\pm$0.14, and 13.3$\pm$0.3 Hz and
no broken power law needed (a weak, single power law at low
frequencies was used instead). This is reminiscent of the complex of
harmonically related QPOs that was found during later observations
(see Section \ref{type-b_sec}). A small frequency shift between
different observations would in this already be enough to smooth out
all but the most significant peaks.

On MJD 51145 the source had moved considerably down branch I, towards
the DBB curve in the CD (HC$\sim$0.01). The count rate had dropped
from $\sim$530 counts s$^{-1}$, in the previous observations, to
$\sim$340 counts s$^{-1}$. The power spectrum did not show any QPO and
was fitted with a power law with a strength of less than 1\% ($\alpha$
fixed to 1). Although the count rate actually went down, the  weak
noise, the absence of QPOs, and the softer X-ray spectrum indicate
that the source had changed from the IS to the HS.

On MJD 51147 the location in the CD was close to that of the first
three observations, on MJD 51139--51143, and the count rate had
increased again to $\sim$450 counts s$^{-1}$. The power spectrum
looked similar to that of MJDs 51139--51143, but the more complex
shape of the noise made it necessary to use a fit function comprised
of a power law, a power law with an exponential cutoff (for the low
frequency noise), and a Lorentzian (for a QPO around 7 Hz). The
power law component had an rms amplitude of 1.31$^{+0.42}_{-0.26}$\%
and an index of 1.2$^{+0.4}_{-0.3}$. The cutoff power law had an rms
amplitude of 2.6$\pm$0.2\% (0.01--1 Hz), $\nu_{cutoff}=7\pm1$ Hz, and
an index of 0.0$\pm$0.1. The QPO at 7.4$\pm$0.3 Hz had a FWHM of
2.3$^{+0.7}_{-0.5}$ Hz and an rms amplitude of 2.6$^{+0.4}_{-0.3}$\%.
Like in the MJD 51139--51143 observations, the power spectrum showed a
strong energy dependence. The source had probably returned to the IS.

On MJD 51150 the source was very close to the MJD 51145 observation in
the CD, although the count rate was somewhat higher ($\sim$420 counts
s$^{-1}$). The power spectrum showed no QPOs, and the power law noise
was weak (less than 1\%), which is typical for a HS.
  
After MJD 51150 the source moved closer to the DBB curve, along branch
I. From MJDs 51152 to 51178 (HC$\sim$0, SC$\sim$0.5--0.13) the
individual power spectra could be described by a single power law with
an index of $\sim$1 and a strength that increased from $\sim$0.5\% to
$\sim$2\% rms. During this period the count rate increased from $\sim$470
to $\sim$7950 counts s$^{-1}$.

\begin{figure}[t]
\centerline{\psfig{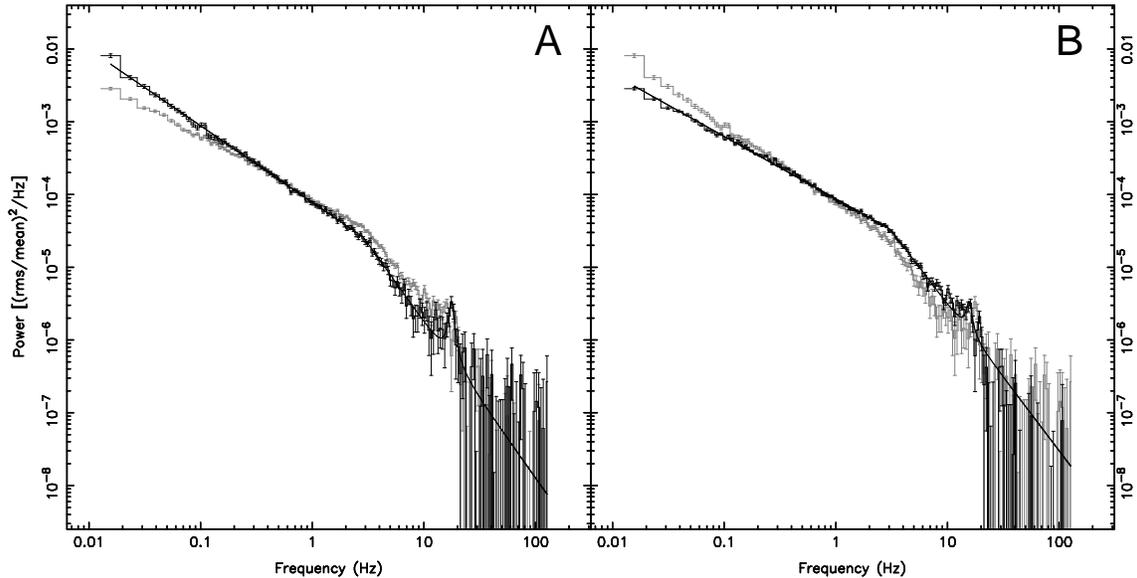}}
\caption{\scshape \small A comparison of the combined flare and interflare power
spectra. Both (a) and (b) show the flare and interflare power
spectra. In (a) the fitted flare power spectrum (black) is shown, with
the interflare power spectrum (gray) for comparison, and in (b) the
fitted interflare power spectrum (black) is shown, with the flare
power spectrum (gray) for comparison.The solid black lines in (a) and
(b) are the best fits with a broken power law and a QPO (see Table
\ref{flare_tab} for fit results). The power spectra in this figure are
rms normalized, and the Poisson level was
subtracted\label{fl-int_pow_fig}}
\end{figure}

\begin{table}[h]
\begin{center}
\begin{tabular}{ccc}

                       & Flares                    & Interflares             \\
\hline 
\hline
$\nu_b$ (Hz)            & 2.87$^{+0.10}_{-0.16}$    & 2.84$^{+0.04}_{-0.06}$ \\
$\alpha_1$              & 1.05$\pm$0.01             & 0.85$\pm$0.01          \\
$\alpha_2$              & 2.1$\pm$0.01              & 1.72$\pm$0.03          \\
0.01--1 Hz rms (\%)     & 2.012$\pm$0.011           & 1.716$\pm$0.005        \\
1--10 Hz rms (\%)       & 1.141$\pm$0.008           & 1.353$\pm$0.005        \\
QPO Frequency (Hz)      & 17.87$\pm$0.17            & 15.6$^{+0.2}_{-0.3}$   \\
QPO FWHM (Hz)           & 1.6$\pm$0.5                & 2.0$^{+1.5}_{-0.8}$   \\
QPO rms (\%)            & 0.30$\pm$0.03             & 0.27$^{+0.06}_{-0.04}$ \\
\hline
\end{tabular}
\end{center}
\caption{\scshape \small Power spectral properties of the combined flare and interflare
observations in the 2--60 keV band.\label{flare_tab}}
\end{table}

\subsection{The broad maximum and the flares}
During the broad maximum of the source five flares were observed (see
Figure \ref{lc_fig}). We compared the averaged power spectrum of these
flares with that of the parts between the flares (hereafter
`interflares'). For the interflare observations we took all
observations between MJDs 51170 and 51237 that were neither in a
flare, nor within one observation from a flare. Although their hard
color (Fig. \ref{lc_fig}h) and 0.01--0.1 Hz noise (Fig.
\ref{noise_curve_fig}a) showed behavior similar to that of the flares,
the observations on MJD 51220 were not included in either category,
since they did not show up as a flare in the light curves and the CD.

\begin{figure}[t]
\centerline{\psfig{figure=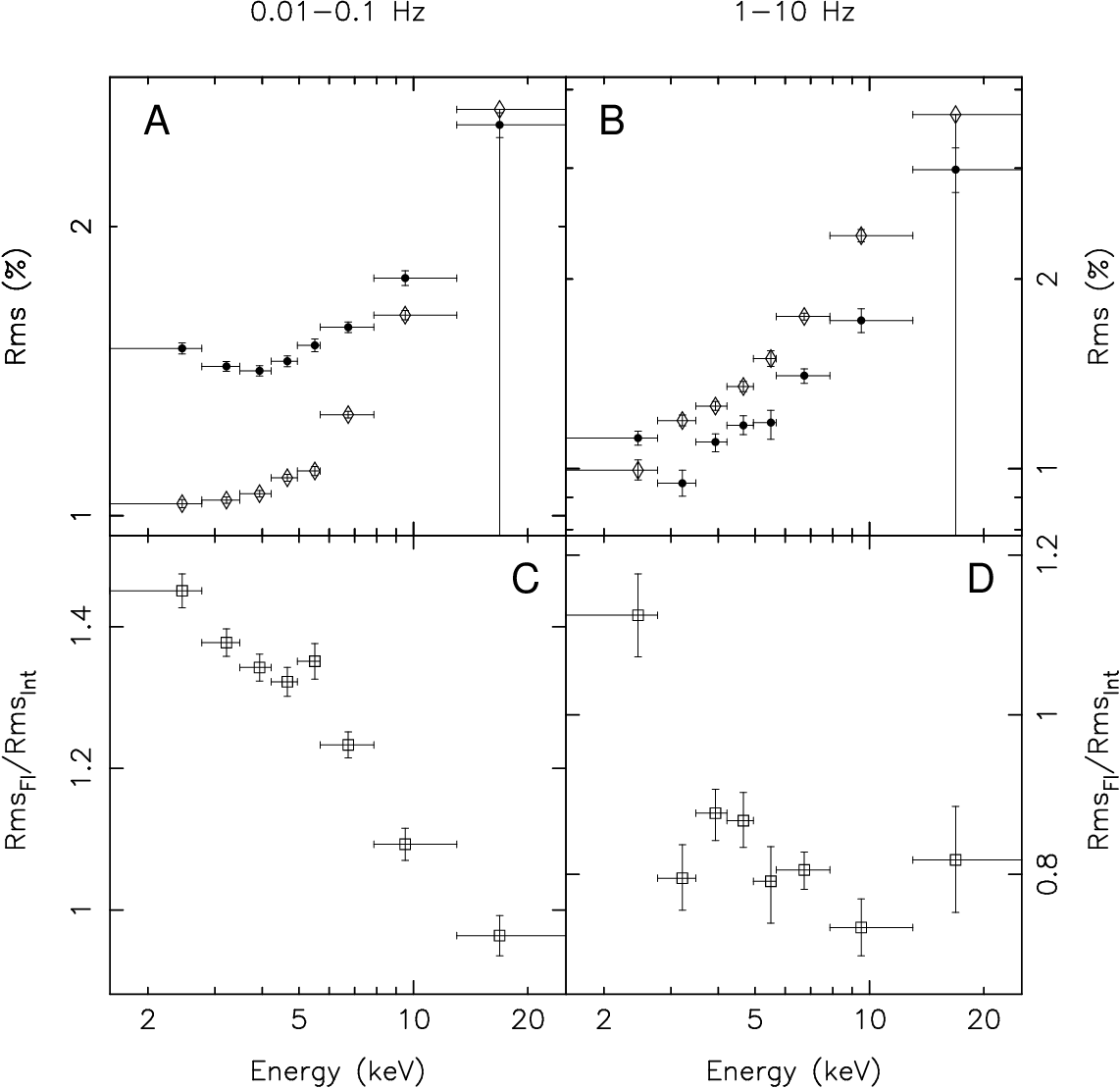,width=7cm}}
\caption{\scshape \small The energy dependence of the 0.01--0.1 Hz (a) and 1--10 Hz
(b) noise components in the flares (bullets) and interflares
(diamonds). The ratios of these energy spectra are shown in (c) and
(d). \label{fl-int_energy_fig}}
\centerline{\psfig{figure=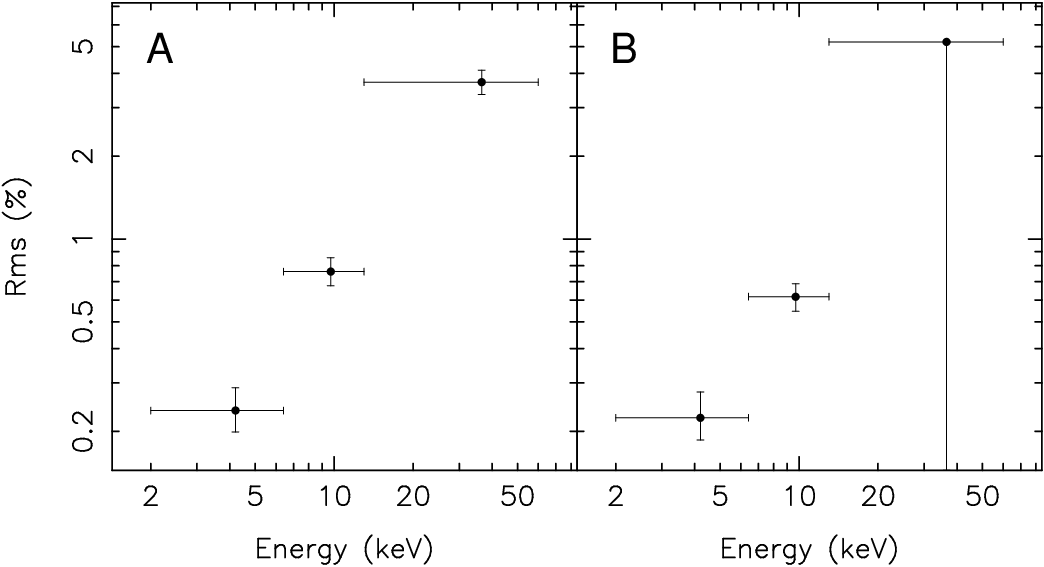,width=7cm}}
\caption{\scshape \small The energy spectra of the 18 Hz QPO in the flares (a) and the
16 Hz QPO in the interflares (b). The highest energy point in (b) is an
upper limit.\label{17_energy_fig}}
\end{figure}

Figure \ref{fl-int_pow_fig} shows the 1/128--128 Hz power spectra
(2--60 keV) of the combined flare observations and the combined
interflare observations. For reasons explained in Section
\ref{obs_sec}, we did not use a single power law to fit the noise (as
we did for the individual observations). Instead, we used a broken
power law, with a Lorentzian for a QPO around 15--18 Hz.  The power
spectral fit parameters for the flare and interflare observations
are given in Table \ref{flare_tab}. Both power spectra show a clear
break around 3 Hz. The noise component in the power spectrum of the
flares is steeper than that in the interflare power spectrum, both
below and above the break. The rms normalized power spectra of the
flares and interflares cross each other around 1 Hz (see Figure
\ref{fl-int_pow_fig}), with the 0.01--1 Hz noise being stronger in the
flares, and the 1--10 Hz noise being stronger in the interflares (see
Table \ref{flare_tab}).  

Figure \ref{noise_curve_fig}a shows the
strength of the 0.01--0.1 Hz noise as a function of time.  When
comparing this figure with Figure \ref{lc_fig}, it is evident that
increases in the strength of the 0.01--0.1 Hz noise occurred at the
times of the hard flares. Note that we used the 0.01--0.1 Hz noise
instead of the 0.01--1 Hz noise, since the effect is more pronounced
in the 0.01--0.1 Hz range. Figure \ref{fl-int_energy_fig} shows the
energy dependence of the 0.01--0.1 Hz (Fig. \ref{fl-int_energy_fig}a)
and 1--10 Hz (Fig. \ref{fl-int_energy_fig}b) noise, for both the
flares (bullets) and the interflares (diamonds). From this figure it
is apparent that the fractional rms energy spectrum of the 0.01--0.1
Hz noise in the flares was softer than that in the interflares
(Fig. \ref{fl-int_energy_fig}c), as opposed to the spectrum of the
source itself, which was harder in the flares than in the interflares
(Fig. \ref{lc_fig}g,h). Apart from it being stronger in the
interflares below 3 keV, the 1--10 Hz noise showed (Fig.
\ref{fl-int_energy_fig}d) no clear spectral change between the flares
and interflares. Although the detections of the noise are very
significant, we note that the amplitudes are compatible with the HS
observations of other sources. The large amount of data, and the
relatively high count rates of XTE J1550--564 made it possible to
study the HS power spectra in much higher detail than was possible in
other sources before.

Apart from the difference in the noise, some of the individual power
spectra during the flares also showed a QPO around 18 Hz. The only
flare in which the QPO was not significantly detected was flare 3, the
softest flare.  Figure \ref{fl-int_pow_fig}a shows the 2--60 keV power
spectrum of the combined flares (including flare 3), with the QPO at a
frequency of 17.87$\pm$0.17 Hz.  The energy spectrum of the QPO in the
flares is shown in Figure \ref{17_energy_fig}a. In the highest energy
band (13.1--60 keV) the probable second harmonic of the QPO was
detected at 35.5$\pm$2.0 Hz, with a FWHM of 10$^{+4}_{-3}$ Hz and an rms
amplitude of 4.6$^{+1.2}_{-0.6}$\%. In the two lower bands only upper
limits could be determined to the rms amplitude of the harmonic:
0.17\% (2--6.5 keV) and 0.8\% (6.5--13.1 keV).  We also searched for a
QPO around 18 Hz in the combined interflare power spectra; a QPO was
found at 15.6$^{+0.2}_{-0.3}$ Hz. Its energy spectrum is shown in
Figure \ref{17_energy_fig}b. The energy spectra of the 17.87 Hz QPO,
its harmonic, and the 15.6 Hz QPO are consistent with each other.

\subsection{Branch II - The Very High State}\label{vhs}

\begin{figure}[t]
\centerline{\psfig{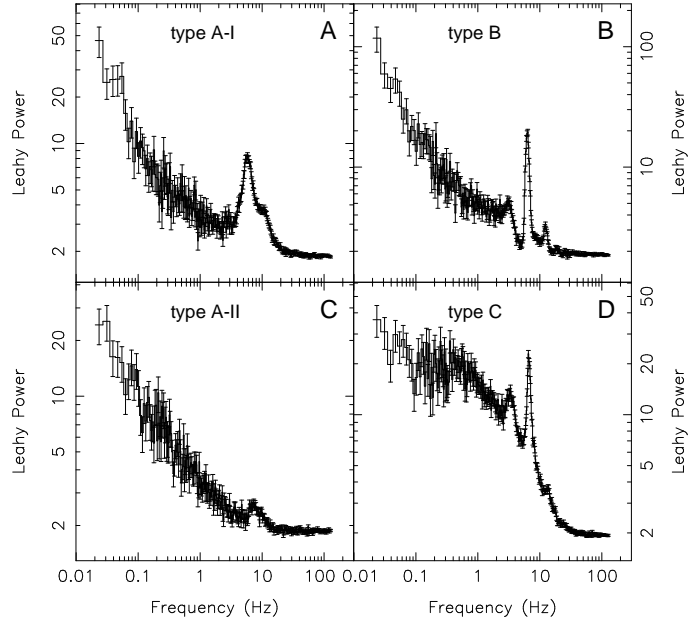}}
\caption{\scshape \small Four types of power spectra that were observed in the MJD
51241--51259 VHS branch (branch II). Type A on MJD 51241 and MJD 51244
(a and c), type B on MJD 51245 (b), and type C on MJD 51250
(d). Poisson level was not subtracted.
\label{qpo-types_fig}}
\end{figure}

\begin{figure}[t]
\centerline{\psfig{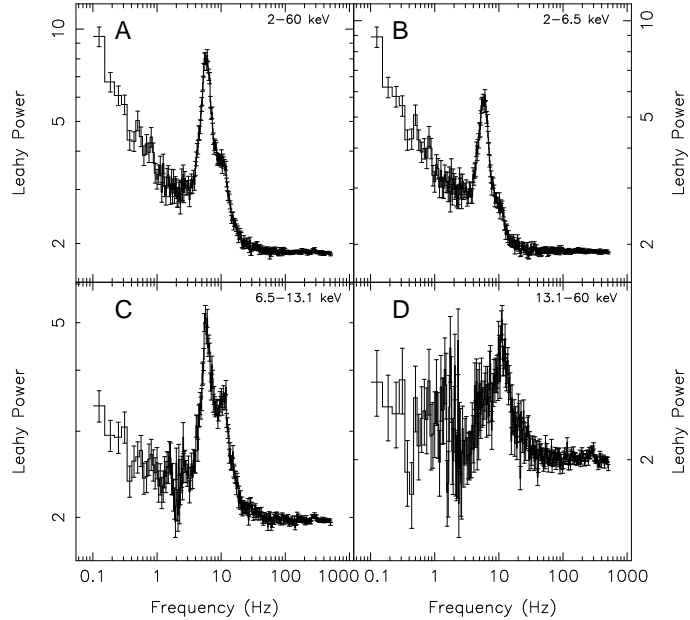}}
\caption{\scshape \small The power spectrum of MJD 51241 in four energy bands. The
$\sim$284 Hz QPO can be seen in the highest energy band. Poisson level
was not subtracted.\label{obs-50_fig}}
\end{figure}

As mentioned in Section \ref{lc-cd_sec}, the source did not return to
the DBB curve after the fifth flare. On MJD 51237 (HC$\sim$0.001,
SC$\sim$0.13) the source reached the position closest to the DBB
curve. After that, on MJDs 51239 and 51240, it started to move away
from the DBB curve, along branch II. On both these days the 18 Hz QPO
was found again. On MJD 51241 the source had moved further away from
the DBB curve than during the previous flares: HC$\sim$0.008,
SC$\sim$0.24. The power spectrum of this observation (see Figures
\ref{qpo-types_fig}a, \ref{obs-50_fig}, and \ref{high-freq_fig}) was
rather different from those seen during the HS and the flares. A broad
peak around 6 Hz was found, and also a peak around 280 Hz. The
presence of the 6 Hz and 280 Hz peaks, the reappearance of the hard
component in the energy spectrum, and the much higher count rate
compared to branch I suggest that the source had entered the VHS, as
was already reported by \citet{howiva1999}. This VHS lasted from MJD
51241 until MJD 51259 (the entire branch II in the CD). 

In the power spectra of nearly all the VHS observations one or more
QPOs are present around 6 Hz. Although the frequency of this QPO
varied between 5 and 9 Hz, for reasons of clarity this QPO will be
referred to as `the 6 Hz' QPO. Some observations also show a single
QPO with a frequency between 100 and 300 Hz.  Based on the Q-value
(the QPO frequency divided by the QPO FWHM) of the 6 Hz QPO, and its
harmonic structure, \citet{wihova1999} distinguished two types of VHS
power spectra: one type with a relatively broad ($Q<3$) 6 Hz QPO and
sometimes a harmonic at 12 Hz (type A low-frequency QPOs; see
Sec. \ref{type-a_sec}), and one with a relatively narrow ($Q>6$) 6 Hz
QPO, with harmonics at 3 and 12 Hz (type B low-frequency QPOs; see
Sec. \ref{type-b_sec}). We decided to divide type A into two
subclasses; one in which the 6 Hz QPO is strong (rms $>$ 2\%) and the
harmonic at 12 Hz was detected (type A-I), and one in which the 6 Hz
QPO was weak (rms $<$ 2\%) and no harmonic was detected
(type A-II). In addition to type A and B a third type, type C, was
introduced by \citet{soremu2000b}, which mainly occurred during the
first part of the outburst. Its harmonic structure is similar to that
of type B, but the 6 Hz QPO has a higher Q-value (Q$\ga$10) and its
time lag behavior is different. The only type C observation, found by
\citet{soremu2000b}, during the second part of the outburst (MJD
51250) was already classified as an odd type B observation by
\citet{wihova1999}. Since this observation and the one on MJD 51254
showed odd behavior, compared to the other VHS observations, they will
be discussed separately in Section \ref{type-spec_sec}. Figure
\ref{qpo-types_fig} shows representative 1/128--128 Hz 2--60 keV power
spectra of type A-I (Fig. \ref{qpo-types_fig}a, MJD 51241), A-II
(Fig. \ref{qpo-types_fig}c, MJD 51244), B (Fig. \ref{qpo-types_fig}b,
MJD 51245), and C (Fig. \ref{qpo-types_fig}d, MJD 51250).

In Table \ref{types_tab} the types of all observations on branch II
can be found. In the remainder of this section we describe the three
types of low frequency power spectra and the high frequency QPOs in
more detail.

\begin{table}[h]
\begin{tabular}{llccccccc}

Date       & Type     &   Frequency     &  FWHM &    rms        &     rms         &    rms              &    rms             \\
           &          &                 &       &  {\small(2--60 keV)} & {\small(2--6.5 keV)} &  {\small(6.5--13.1 keV)} &  {\small(13.1--60 keV)} \\
(MJD)      &	       &  (Hz)           & (Hz)  &    (\%)	       &  (\%)           &	   (\%)	&          (\%)       \\
\hline
\hline
51241      & A-I      &   284$\pm$2     &  30$^{+5}_{-4}$  & 1.06$^{+0.14}_{-0.11}$ & $<$0.9     & 2.8$\pm$0.3         & 7.7$^{+0.9}_{-0.8}$\\
51242      & A-I      &   282$\pm$3     &  32$^{+9}_{-5}$  & $<$1.06	         & $<$1.0     & 3.3$^{+0.7}_{-0.5}$ & $<$7	     \\
51244      & A-II     &	          & & & & &					              \\
51245      & B        &   178$\pm$3     & 22$^{+12}_{-7}$  & 1.11$^{+0.19}_{-0.15}$ & 0.94$^{+0.19}_{-0.11}$ & 1.5$^{+0.3}_{-0.2}$ & $<$4   \\
51246      & A-II     & 208$^{+2}_{-3}$ & 16$^{+13}_{-9}$  & 0.76$\pm$0.10 & 0.92$^{+0.21}_{-0.12}$ & $<$1.2 & $<$3.3		 \\
51247      & B        &   187$\pm$5     & 70$\pm$20        & 1.65$^{+0.23}_{-0.18}$ & 1.54$\pm$0.19 & 2.51$^{+0.25}_{-0.10}$ & $<$5	     \\
51248      & B        &	          & & & & &					            \\
51249      & B        &	          & & & & &					              \\
51249      & B        &	          & & & & &					             \\
51250      & C        &   102$\pm$2     & 18$^{+11}_{-6}$  & $<$1.3 & $<$0.9 & 2.8$\pm$0.4 & $<$3.3			      \\
51253      & B        &	          & & & & &					            \\
51254$<$   & C$^1$    &            & & & & & \\
51254$>$   & B$^1$    &   123$\pm$2     & 13$\pm$2         & $<$0.8 & $<$0.7 & 2.4$\pm$0.3 & $<$3.5			      \\
51255.08   & A-I      &	          & & & & &					             \\ 
51255.12   & A-I      &   280$\pm$3     & 40$\pm$10        & 1.22$\pm$0.19 & $<$1.1 & 2.37$\pm$0.31 & 9.5$\pm$1.0		      \\
51257      & A-II$^1$     &	          & & & & &					             \\
51258.1      & A-I    &	          & & & & &					             \\
51258.5      & A-I    &	          & & & & &					             \\
51258.9 & A$^1$     &  	 & & & & &	        \\
51259    & A$^1$     &  	 & & & & &	        \\
\hline
\end{tabular}
\caption{\scshape \small Classification of the VHS (branch II) power spectra. For
those observations where a high frequency QPO was found the frequency,
FWHM, and the rms amplitudes in four energy bands are given. The
frequency and FWHM are taken from the band where the QPO was most
significant.\newline$^1$The type of the low frequency QPOs was
uncertain.\label{types_tab}}
\end{table}

\subsubsection{Type A observations}\label{type-a_sec}
\begin{figure}[t]
\centerline{\psfig{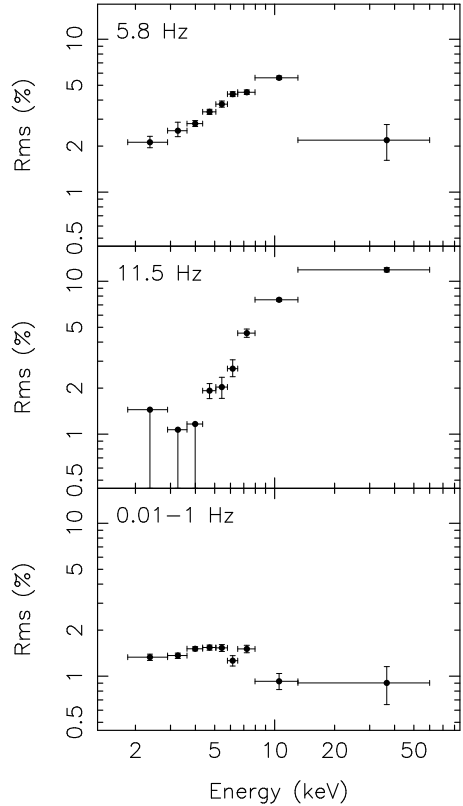}}
\caption{\scshape \small The energy spectrum of the power spectral components of the
MJD 51241 (type A-I) observation. Points whose negative rms error
extends to the bottom edge are upper limits.\label{obs-50_energy_fig}}
\end{figure}

On MJD 51241 (HC$\sim$0.008) the first observation in the very high
state was made. The power spectrum was of type A-I, and it can be
regarded as representative for the other type A-I power spectra. It is
therefore the only type A-I power spectrum that will be discussed in
detail in this paper. Figure \ref{obs-50_fig} shows the power spectrum
of MJD 51241 in four energy bands.  Apart from a QPO around 6 Hz,
there was some excess around 11 Hz, which in the high energy power
spectra showed up as a QPO. In the 13.1--60 keV band, the 6 Hz peak
had almost disappeared. The two peaks seemed to be harmonically
related, but fitting them simultaneously in the 2--60 keV band gave
frequencies that were not consistent with the two being harmonically
related: 5.93$\pm$0.03 Hz and 10.41$\pm$0.13 Hz. However, when
comparing the frequency of the lower frequency QPO in the 2--6.5 keV
band (5.85$\pm$0.03 Hz) with that of the higher frequency QPO in the
13.1--60 keV band (11.52$\pm$0.19 Hz) we found a ratio of
1.97$\pm$0.03, which suggests an harmonic relation (see also
\citet{wihova1999}). The reason that the frequencies differed so much
between the energy bands might be that the chosen fit function was not
appropriate, or that an extra component was present between the two
QPOs. The FWHM of the 5.85 Hz QPO in the 2--6.5 keV band was
2.41$\pm$0.01 Hz, and that of the 11.52 Hz QPO in the 13.1--60 keV
band 7.23$\pm$0.7 Hz. Their rms amplitudes in the 2--60 keV band were
3.00$\pm$0.05\% and 2.64$^{+0.09}_{-0.08}$\%. At low frequencies a
noise component was present. It could be fitted with a single power
law with an index of 0.90$\pm$0.03, and a strength of 1.31$\pm$0.03\%
rms.  Figure \ref{obs-50_energy_fig} shows the photon energy spectra
of the two low frequency QPOs and the noise component. The frequencies
of the low frequency QPOs were not fixed, for reasons explained
above. The 5.8 Hz QPO first increased in strength with energy, but
above 10 keV it dropped by a factor of $\sim$2.5. Its harmonic showed
a strong increase with photon energy, from $\sim$1\% rms in the lowest
energy bands to more than 11\% rms in the highest band. The 0.01--1 Hz
noise had a relatively flat energy spectrum, with strengths between
1\% and 2\% rms, although it became slightly weaker ($\sim$0.9\%)
above 10 keV. Selections were  made on time, color and count rate,
but no significant dependencies were found.

\begin{figure}[t]
\centerline{\psfig{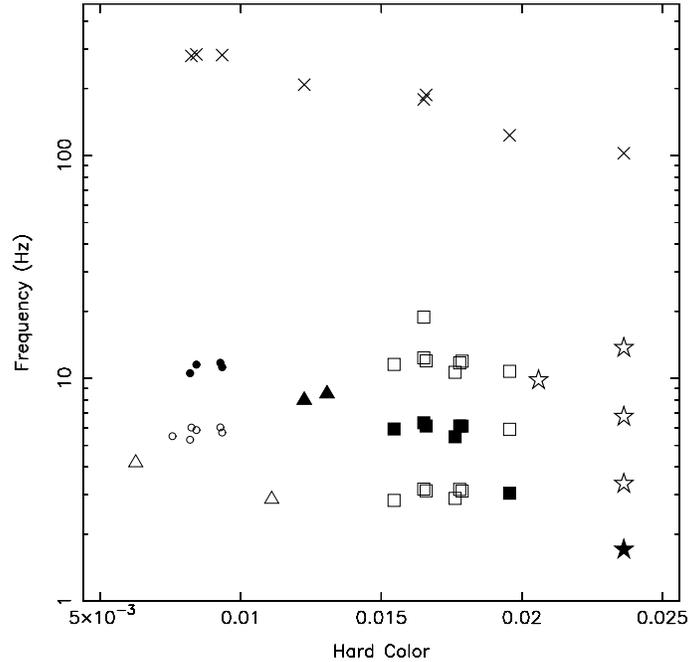}}
\caption{\scshape \small The frequencies of the QPOs in the MJD 51241--51259 (VHS,
branch II) power spectra as a function of hard color. Crosses depict
the high frequency QPOs, circles type A-I, triangles type A-II,
squares type B, and stars type C. Filled symbols represent those QPOs
that according to Figure \ref{low-high-qpo_fig} can be identified with
the same harmonic component. The 2.9 Hz QPO found in the MJD 51259
observations is shown as a triangle (HC$\sim$0.011), although its type
(A-I to A-II) was uncertain. The QPOs of the MJD 51254 observation are
plotted as type B (HC=0.195; after jump) and type C (HC=0.205; before
jump), although their types were also uncertain.\label{qpo-hc_fig}}
\end{figure}

The next observation, on MJD 51242, was located close to  the previous
observation in the CD, and its power spectrum was also very
similar. Again two low frequency QPOs were found, with frequencies of
5.71$\pm$0.01 Hz (2--6.5 keV) and 11.2$\pm$0.3 Hz (13.1--60 keV), FWHM
of 3.1$\pm$0.4 Hz and 8$\pm$1 Hz, and rms amplitudes (2--60 keV) of
2.53$^{+0.11}_{-0.12}$\% and 2.51$^{+0.17}_{-0.14}$\% respectively.

In the power spectrum of MJD 51244 (see Figure \ref{qpo-types_fig}c) a
QPO was found with a frequency of 8.5$\pm$0.3 Hz, a FWHM of
3.8$^{+0.9}_{-0.7}$ Hz, and an rms amplitude of
1.34$^{+0.13}_{-0.12}$\% (2--60 keV). The QPO was considerably weaker
than the 6 Hz QPOs in the two previous observations ($\sim$3\%
and $\sim$2.5\%, 2--60 keV), and no sub- or second harmonics were
found. The energy dependence of the QPO was rather steep, but in the
highest band only an upper limit could be determined: 0.8$\pm$0.2\%
(2--6.5 keV), 2.9$\pm$0.3\% (6.5--13.1 keV), and $<$3.7\% (13.1--60
keV). Although the Q-value is similar to that of the previous two
observations the above characteristics set this observation apart, and
we therefore defined it to be of type A-II. In the CD the observation
was located further along branch II, away from the DBB line and
towards a stronger power law spectral component (HC$\sim$0.13).

The next type A observations occurred on MJD 51246, after a type B
observation on MJD 51245 (see Sec. \ref{type-b_sec}). In the CD it
was located close to the MJD 51244 observation. A QPO was found at
7.72$\pm$0.13 Hz, with a FWHM of 3.1$\pm$0.4 Hz. Since the QPO was rather
weak (1.20$\pm$0.06\% rms) and no harmonics were found, it was
classified as type A-II.

During MJDs 51247--51254 the source moved further up branch II in the
CD, and the power spectra only showed type B and C QPOs (see
Secs. \ref{type-b_sec} and \ref{type-spec_sec}). Type A QPOs
reappeared on MJD 51255, when the source returned to a location in the
CD close to the other type A observations (see Figure \ref{cd_fig}:
HC$\sim$0.01, SC$\sim$0.2).  From MJD 51255 to 51258.5 the source
showed both type A-I and A--II QPOs, with similar properties as those
in the beginning of the VHS. The power spectrum of MJD 51258.9
showed a broad feature around 6 Hz that had a width larger than 6 Hz,
and the power spectrum of MJD 51259 showed a QPO at $\sim$3 Hz with a
FWHM of $\sim$1 Hz, and a broad (FWHM$\sim$12 Hz) peak around 9 Hz. We
classified these two power spectra as type A, but it was not clear of
what sub-type they are; both broad peaks might have been unresolved
pairs of harmonics.

Figure \ref{qpo-hc_fig} shows the frequency of all VHS (branch II)
QPOs as a function of the hard color, which is a good measure of the
position along branch II, as can be seen from Figure \ref{cd_fig}.
The frequencies of the broad peaks in the MJD 51258.9 and MJD 51259
power spectra are not included. Note that type A-II (triangles)
observations were located both at lower and at higher hard colors with
respect to the type A-I (circles) observations in Figure
\ref{qpo-hc_fig}.

\begin{figure}[t]
\centerline{\psfig{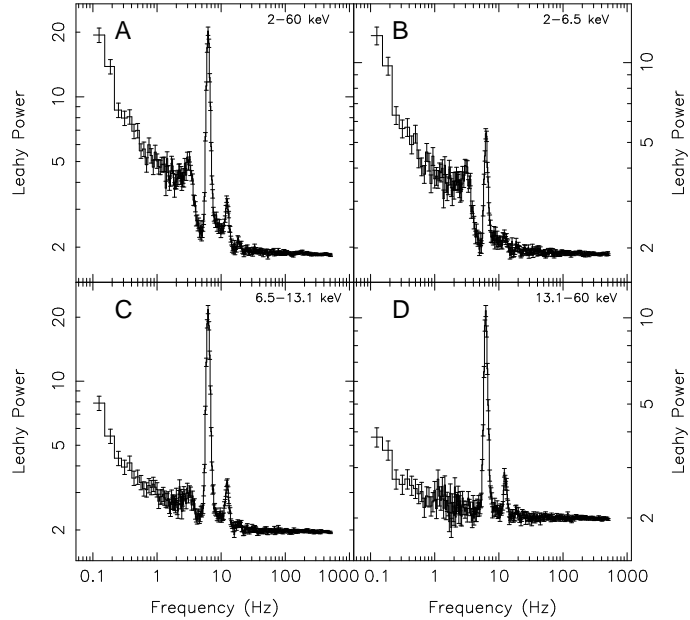}}
\caption{\scshape \small The power spectrum of MJD 51245 in four energy bands. Poisson
level was not subtracted.\label{obs-53_fig}}
\end{figure}

\begin{table}[thb]
\begin{center}
\begin{tabular}{ccc}
 Frequency & FWHM & Rms Amplitude \\
 (Hz)      &  (Hz)   & (\%)          \\
\hline
\hline
3.15$\pm$0.03       & 1.3$\pm$0.1            & 1.54$^{+0.06}_{-0.19}$ \\
6.319$\pm$0.008     & 0.88$\pm$0.02          & 3.35$\pm$0.04         \\
7.9$\pm$0.2         & 3.4$\pm$0.3            & 1.39$\pm$0.09           \\
12.35$\pm$0.06      & 2.27$\pm$0.14          & 1.54$\pm$0.04         \\
18.85$\pm$0.25      & 2.6$^{+0.6}_{-0.5}$    & 0.60$\pm$0.07         \\
\hline
\end{tabular}
\end{center}
\caption{\scshape \small Fit values for the Gaussian peaks in the power spectrum of
MJD 51245. Note that the FWHM in this case refers to the width of a Gaussian and
not to that of a Lorentzian. \label{obs_53_tab}}
\end{table}

\subsubsection{Type B observations}\label{type-b_sec}
\begin{figure}[t]
\centerline{\psfig{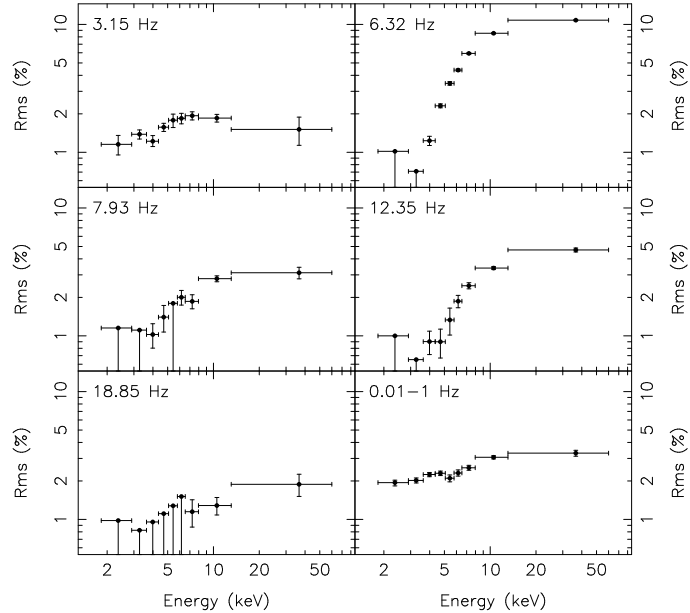}}
\caption{\scshape \small The energy spectrum of the power spectral components of the
MJD 51245 (type B) observation. Points whose negative rms error extends 
to the bottom edge are upper limits.\label{obs-53_energy_fig}}
\end{figure}

The first type B observation occurred on MJD 51245. The source had
moved further up branch II (HC$\sim$0.016), compared to the previous
(type A) observations. Figure \ref{obs-53_fig} shows the power
spectrum of this observation in four energy bands. A very sharp QPO
was present around 6 Hz, with harmonics around 12 and 18 Hz, and a
sub-harmonic around 3 Hz. There were also indications for a peak
around 24 Hz, but our fits showed it was not significant.  Fitting the
power spectrum with a power law and Lorentzians (six in total) gave a
poor result ($\chi^2_{red}=2.6$, $d.o.f.=230$). We tried using a power
law and Gaussian functions (again six) instead, which improved the
quality of the fit ($\chi^2_{red}=1.4$, $d.o.f.=230$, see also
\citet{wihova1999}). The two extra peaks, at $\sim$0.2 Hz and at
$\sim$1.25 times the frequency of the 6 Hz QPO, were added to the fit
function to account for a low frequency component, and for the
shoulder of the 6 Hz QPO, respectively. The fit results (2--60 keV)
for the four QPOs and the shoulder component are given in Table
\ref{obs_53_tab}. The QPOs are not perfectly related harmonically,
most likely because the fit function did not describe the data well
enough. The 0.01--1 Hz noise was fitted with a single power law, with
$\alpha=1.8\pm0.1$ and an rms amplitude of 5.6$^{+0.3}_{-0.2}$\%.  The
photon energy spectra of the various power spectral components are
shown in Figure \ref{obs-53_energy_fig}. Except for the 3 Hz QPO, and
the noise component, all QPOs showed a considerable increase in
strength with photon energy. The 0.01--1 Hz noise only showed a weak
increase, and the 3 Hz QPO behaved similar to the 6 Hz QPO in the type
A-I power spectra, in that it seemed to become weaker above 10
keV. Selections were made on color, time and count rate. It was found
that the harmonic at 18 Hz was more significant at low hard colors.

Other type B power spectra were found between MJD 51247 and MJD 51253,
with QPOs that were similar to those on MJD 51245. Their 6 Hz QPOs had
frequencies between 5.3 Hz and 6.1 Hz, rms amplitudes between 3.3\%
and 3.4\%, and Q-values between 6.2 and 7.3. The 3 Hz QPOs had rms
amplitudes between 1\% and 2.3\% and showed a weak trend of an
increase with hard color. The MJD 51245 observation remained the only
one in which the harmonic around 18 Hz was significantly
detected. Figure \ref{qpo-hc_fig} shows the frequencies of the type B
QPOs (squares) as a function of the hard color.

\begin{figure}[t]
\centerline{\psfig{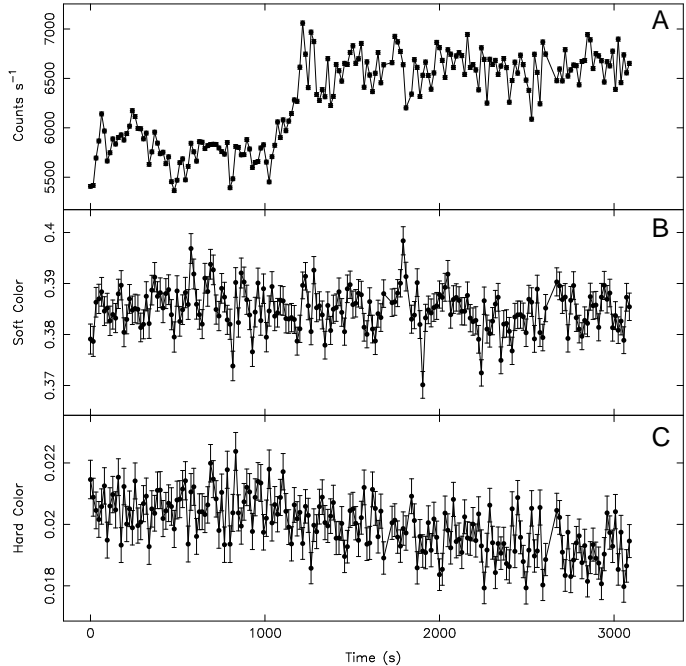}}
\caption{\scshape \small The 2--60 keV light curve (a), soft color curve (b), and hard
color curve (c) of the MJD 51254 observation. The transition (around
t=1100 s) can clearly be seen in the light curve. The soft color (b)
does not change significantly, whereas the hard color (c) starts to
decrease slowly after t=1100 s.\label{transition_fig}}
\end{figure}

\subsubsection{Special cases: Strong noise on MJD 51250 and the transition on MJD 51254}\label{type-spec_sec}
Although, based on the Q-value ($\sim$8) of the 6 Hz QPO and the
harmonic content, the power spectrum of the MJD 51250 observation was
classified as type B by \citet{wihova1999}, they also found that the
time lag behavior of this observation was quite different from that of
the other type B observations. Based on this time lag behavior the
observation was classified as type C by \citet{soremu2000b}, making it
one of only two (see below for the second) type C observations during
the second part of the outburst. More deviations from the type B power
spectra were found; in addition to the power law noise (1.6$\pm$0.2
\%) a strong noise component was present at 0.1--1 Hz (see Figure
\ref{qpo-types_fig}d; also \citet{wihova1999}). This noise component,
which we fitted using a zero-centered Lorentzian with a width of
$\sim$3 Hz, was present in all the energy bands, with rms amplitudes
of 13.1$\pm$0.3\% (2--60 keV), 3.3$\pm$1.2\% (2--6.5 keV),
18.6$\pm$0.5\% (6.5--13.1 keV), and 28.0$\pm$0.8\% (13.1--60
keV). Four low frequency QPOs were found in the 2-60 keV band at (rms
amplitudes in brackets) 1.7$\pm$0.6 Hz (2.74$^{+0.77}_{-0.45}$\%),
3.35$\pm$0.03 Hz (5.8$\pm$2.0\%), 6.68$\pm$0.15 Hz (6.7$\pm$0.2\%),
and 13.68$\pm$0.15 Hz (3.2$^{+0.3}_{-0.2}$\%).  When comparing these
numbers with those of the type B observations, it shows that the rms
amplitudes of the 3 Hz and 6 Hz QPOs are, respectively, a factor
$\sim$2.5--6 and $\sim$2 higher than in the type B observations. The
QPO frequencies are shown as stars in Figure \ref{qpo-hc_fig}. In the
ASM and PCA light curves the MJD 51250 observation is clearly visible
as a dip (see Figures \ref{asm_fig} and \ref{lc_fig}), with a count
rate of only $\sim$5150 counts s$^{-1}$, compared to $\sim$8200 counts
s$^{-1}$ on MJD 51249 and $\sim$7025 counts s$^{-1}$ on MJD
51253. This dip is strongest at low energies, causing a hardening of
the spectrum a (see Figure \ref{lc_fig}).

Figure \ref{transition_fig} shows the 2--60 keV light curve and the
color curves of the MJD 51254 observation. Clearly visible is the jump
in count rate that occurred around 1100 s after the start of the
observation. The soft color seemed to be unaffected by this change,
and though the hard color showed a small change ($\sim$10\%) it was
more gradual than the change in count rate. It should be noted that
the observed transition is not related to the temporary gain change
that was applied to the PCA later during this observation (around
t=3100 s; not shown here).

\begin{figure}[t]
\centerline{\psfig{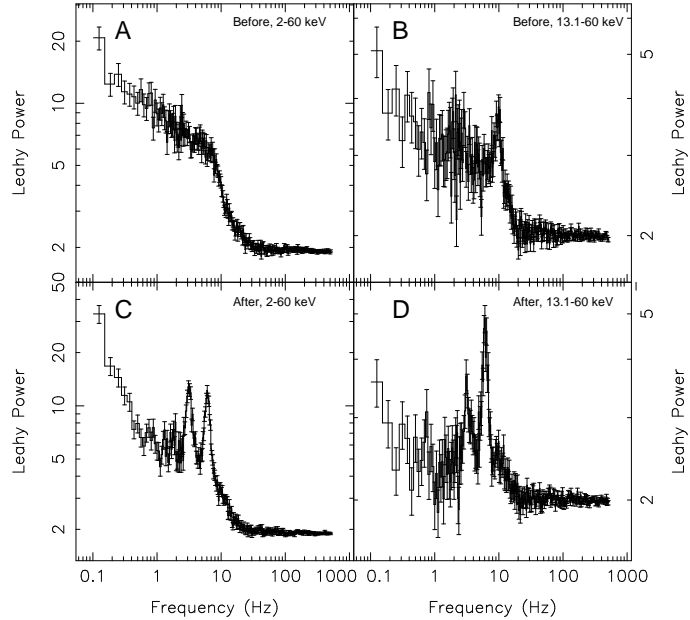}}
\caption{\scshape \small Power spectra of the MJD 51254 observation in two energy
bands, before and after the transition. Poisson level was not
subtracted.\label{obs-58_fig}}
\end{figure}

\begin{figure}[t]
\centerline{\psfig{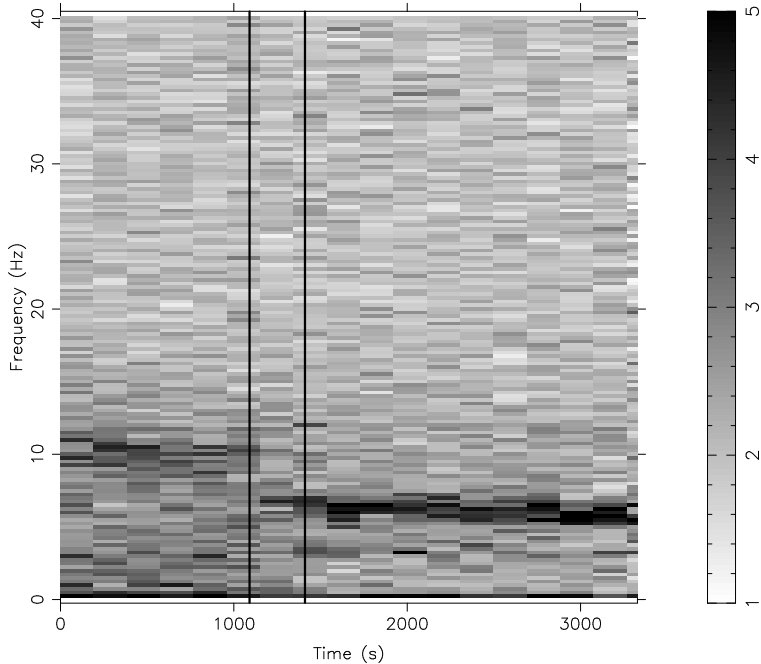}}
\caption{\scshape \small Dynamic power spectrum of the MJD 51254 observation in the
13.1--60 keV band. The first vertical line depicts the approximate
time of transition, the second vertical represents a $\sim$300 s data
gap that was only present in the high time resolution data (and
therefore not visible in Figure \ref{transition_fig}). The time
resolution is 3$\times$64=96 s, and the frequency resolution
32$\times$1/64=0.5 Hz .\label{dyn-trans_fig}}
\end{figure}

Figure \ref{obs-58_fig} shows the power spectra from before (0--1000
s) and after (1500--3000 s) the jump in the 2--60 and 13.1--60 keV
bands.  The 2--60 keV power spectrum before the jump showed a broad
noise component around a few Hz, that was fitted with a power law with
an exponential cutoff. It had a strength (1--100 Hz) of 6.7$\pm$0.1\%
rms, a power law index ($\alpha$) of $-$0.7$\pm$0.1, and a cutoff
frequency of 3.9$\pm$0.3 Hz. In the 13.1--60 keV band the strength of
this component was 12.5$^{+0.9}_{-0.7}$\% rms. In that same band we
found a QPO at 9.8$\pm$0.1 Hz, with an rms amplitude of
8.6$^{+0.6}_{-0.5}$\% and a FWHM of 3.2$\pm$0.5 Hz.  The post-jump
2--60 keV power spectrum showed a similar noise component as before
the jump, though somewhat weaker (4.7$\pm$0.1\% rms), with two QPOs
superimposed on it, at 3.17$\pm$0.02 Hz and 6.14$\pm$0.03 Hz. These
QPOs had rms amplitudes and FWHM of 2.3$\pm$0.1\% and 0.71$\pm$0.06 Hz
(3.17 Hz QPO), and 3.08$\pm$0.08\% and 1.23$\pm$0.06 Hz (6.14 Hz QPO),
respectively. In the post jump 2--6.5 keV power spectrum an additional
QPO at 1.77$\pm$0.04 Hz was found (3.7$\sigma$) when the high count
rate part was selected. Both before and after the transition power law
noise was present: respectively, 3.4$\pm$0.2\% rms with
$\alpha=0.92\pm0.02$ (before) and 3.8$\pm$0.1\% rms with
$\alpha=1.04\pm0.03$ (after).

The fast transition in the power spectrum can be seen in Figure
\ref{dyn-trans_fig}, which shows the dynamical power spectrum in the
13.1--60 keV band. The time scale for the change in the power spectrum
is similar to that of the transition in the 2--60 keV light curve. It
was not possible to track the QPO across the transition since it
became weaker during the transition.  

\citet{wihova1999} and \citet{soremu2000b} classified the power
spectrum after the jump as type B. Indeed, the strengths of the 3 and
6 Hz QPOs were consistent with those in the other type B
observations. On the other hand, the Q-value of the 6 Hz QPO was only
5, and the 5\% rms noise component under the QPOs was not seen in
other type B observations. The type of the power spectrum before the
jump is not clear either.  The hardness of that part of the
observation suggests type B or C, but the Q-value of the 9.8 Hz QPO
was only $\sim$3. The strength of that QPO was lower than that of the
type B and C 6 Hz QPOs in the same energy band ($\sim$11\% rms), but
higher than that of the type B and C 12 Hz QPOs (5--6\% rms). Since
the power spectrum showed a strong noise component, and the 2--60 keV
count rate was lower than that of the type B part it was most likely
of type C. The QPO frequencies of both parts are shown in Figure
\ref{qpo-hc_fig} (the part before [HC=0.205] as type C, the part after
[HC=0.195] as type B).

The exceptional cases of low frequency QPOs presented in this section
clearly demonstrate that the A, B (and C) classification. which works
well for the majority of the observations, is not able to
unambiguously describe all of them.The characteristics on which this
classification is based (Q-value, harmonic content, time lags) show
strong correlations, but deviations from the usual correlations do
occur.

\begin{figure}[t]
\centerline{\psfig{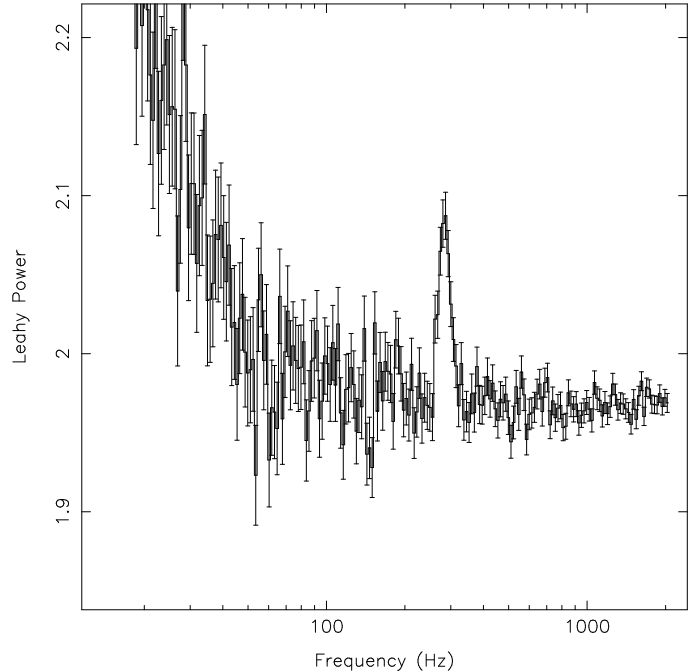}}
\caption{\scshape \small The 13.1--60 keV power spectrum of the MJD 51241 observation,
showing a QPO at 284 Hz. Poisson level was not
subtracted.\label{high-freq_fig}}
\end{figure}

\subsubsection{High frequency QPOs}\label{type-high_sec}
\begin{figure}[t]
\centerline{\psfig{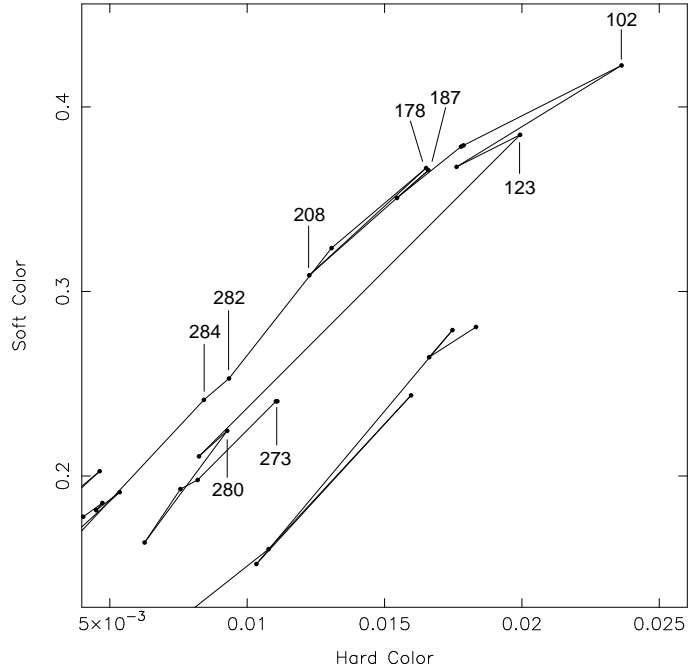}}
\caption{\scshape \small A color-color diagram showing branch II. Observations in
which high frequency QPOs were found are marked with the frequency (Hz) of
the QPO.\label{cd-khz_fig}}
\end{figure}

During several observations in the VHS (branch II), QPOs were found
with frequencies between 100 and 300 Hz. An example can be seen in
Figure \ref{high-freq_fig}, which shows the 284 Hz QPO found in the
power spectrum of MJD 51241. The frequencies of the high frequency
QPOs ($\nu_{HF}$) are given in Table \ref{types_tab}, and the
locations in the CD of the observations in which they were found are
indicated in Figure \ref{cd-khz_fig}. Note that we only report QPOs
whose single-trial significance exceeds 3$\sigma$. It can be seen from
Figure \ref{cd-khz_fig} that $\nu_{HF}$ is related to the location in
the CD. It decreased from 284 Hz to 102 Hz as the source moved up
branch II, and increased again to 280 when it moved down this
branch. Figure \ref{qpo-hc_fig} more clearly shows that $\nu_{HF}$
decreased as the hard color increased.  In the high energy bands, the
QPOs tended to be stronger when they had a frequency around 280 Hz, as
can be seen from Table \ref{types_tab}. The Q-values of the QPOs were
not related to those of the low frequency QPOs, and had values between
5.6 and 13.

We measured time lags for the $\sim$282 Hz QPO in the combined Fourier
spectra of the observations on MJDs 51241, 51242, and 51255. Lags were
measured between three energy bands, in the frequency range 272--292
Hz. All lags were consistent with being zero: 0.00$\pm$0.11 ms (2--6.5
keV and 6.5--13.1 keV), $-$0.08$\pm$0.13 ms (2--6.5 keV and 13.1--60
keV), and $-$0.08$\pm$0.04 ms (6.5--13.1 keV and 13.1--60 keV), where
a positive number means that the photons in the second band lag those
in the first one. A time lag analysis of the low frequency QPOs in the
VHS (branch II) can be found in \citet{wihova1999},
\citet{soremu2000b}, and \citet{cuzhch2000}.

\begin{figure}[t]
\centerline{\psfig{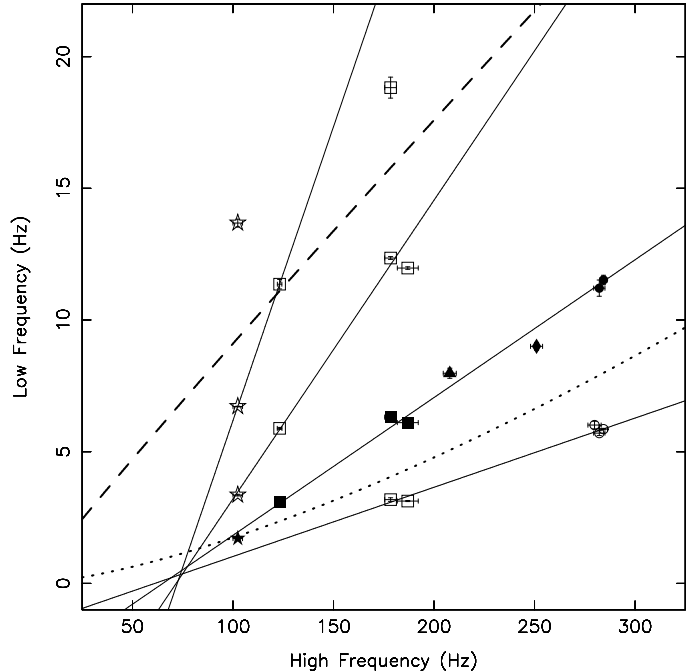}}
\caption{\scshape \small Frequency of the low frequency QPOs as a function of the
frequency of the high frequency QPOs, for those observations where
they could be measured simultaneously. Similar symbols have been used
as in Figure \ref{qpo-hc_fig}, based on the type of the low frequency
QPOs. The four solid lines are the best linear fits to the data. The
two symbols that are not fitted by a solid line are harmonics that
were only observed once. They are consistent with being a 16th
harmonic (highest star at 102 Hz) and a 6th harmonic (highest square
at 178 Hz). Symbols on the second solid line are filled, and have been
used to identify the filled symbols in Figure \ref{qpo-hc_fig}. The
dashed lines represent the relations found by \citet{psbeva1999} for
the Z sources, and the dotted line the relation found by
\citet{di2000a} for 4U 1728--34 (see Section \ref{pow-spec_sec}). The
diamond shows the values for the MJD 51270/51271 observation.
\label{low-high-qpo_fig}}
\end{figure}

Figure \ref{low-high-qpo_fig} shows the frequency of several low
frequency QPOs ($\nu_{LF}$) plotted against $\nu_{HF}$, for those
observations where they were detected simultaneously. We also included
the values for the high frequency QPO that was observed on branch III
(represented by the diamond; see Section \ref{decay_sec}). The 123 Hz
QPO on MJD 51254 was only found in the data taken after the count rate
jump (Section \ref{type-spec_sec}).  This data included a $\sim$600 s
interval (with different PCA gain settings) that was not used for the
analysis of the low frequency QPOs in Figure \ref{qpo-hc_fig}.  The
low frequency QPOs plotted at $\nu_{HF}=123$ Hz in Figure
\ref{low-high-qpo_fig} therefore have a slightly different frequency
than those of the same observation in Figure \ref{qpo-hc_fig}. Four
lines could be fitted to the data points, with the first, third and
fourth line having slopes that were, respectively, 0.50$\pm$0.02,
2.17$\pm$0.10, and 4.3$\pm$0.5 times the slope of the second
line. This is consistent with the four lines representing the
fundamental and second, fourth and eighth harmonics.  Note that the
four lines do not pass through the origin and cross each other around
$\nu_{LF}=0$ Hz and $\nu_{HF}=75$ Hz.  The only two points that were
not fitted by these four lines were the sixth harmonic in the MJD
51245 observation ($\nu_{HF}$=178 Hz) and the sixteenth harmonic in
MJD 51250 observation ($\nu_{HF}$=102 Hz). These components were only
observed once, and therefore no fits could be made. The four lines can
be used to connect the low frequency QPOs in Figure
\ref{qpo-hc_fig}. For example, using the second line in Figure
\ref{low-high-qpo_fig}, it can be seen that the type A-I 10--12 Hz
QPOs are related to the type A-II 8--9 Hz QPOs, the type B $\sim$6 Hz
QPOs, the 3.1 Hz (type B?)  QPO on MJD 51254, and the 1.7 Hz type C
QPO on MJD 51250. The QPOs that lie on the second line in Figure
\ref{low-high-qpo_fig}, and those that based on similarities in the
power spectrum and hard color are expected to, have been represented
by the filled symbols in Figure \ref{qpo-hc_fig}. The filled symbols
show that the frequency of the low frequency QPOs decreases with hard
color, like that of the high frequency QPO.

\subsection{The Decay}\label{decay_sec}
On MJD 51260 (indicated by {\scriptsize BEGIN} in Figure
\ref{cd_fig}b) the power spectrum showed no QPOs; it could be fitted
with a single power law, with a strength of 0.76$\pm$0.05\% rms and an
index of 1.1$\pm$0.01.  This weak noise, the absence of QPOs, and the
relatively soft colors suggest that the source had returned to the HS.

The power spectrum of the next observation (MJD 51261) showed a noise
component with a similar strength (0.56$\pm$0.07\% rms), but also a
QPO at 17.0$^{+0.5}_{-0.3}$ Hz, with a FWHM of 2.5$^{+1.9}_{-1.1}$ and
a rms amplitude of 1.02$^{+0.20}_{-0.16}$\%. Based on the hardness at
which this QPO is found, its frequency, and its FWHM, it may be
related to the 15.6/17.9 Hz QPO that was found in the flare/interflare
observations during MJD 51170--51237 (see section \ref{rise}). In the
next few observations (MJD 51263--51267) no QPO around 17 Hz was
found, and the power spectra could be fitted with single power laws,
with strengths between 0.4\% and 1.2\% rms, typical for HS.

On MJD 51269 the source had started to move up branch III.  Two QPOs
were found in the power spectrum of that observation: at 4.54$\pm$0.15
Hz (1.7$\pm$0.2\% rms, FWHM=1.5$^{+0.5}_{-0.4}$ Hz) and 9.6$\pm$0.6 Hz
(2.0$^{+0.4}_{-0.3}$\% rms, FWHM=4.5$^{+2.4}_{-1.5}$ Hz). The noise at low
frequencies was fitted with a single power law, with a strength of
1.32$\pm$0.07\% rms and an index of 0.8$\pm$0.1. Based on their Q-values, 
 the QPOs are  either of type A-I or A-II; the strength of the QPOs (and
their frequency) suggests type A-II, whereas the detection of an
harmonic suggest type A-I (see section \ref{type-a_sec}).

\begin{figure}[t]
\centerline{\psfig{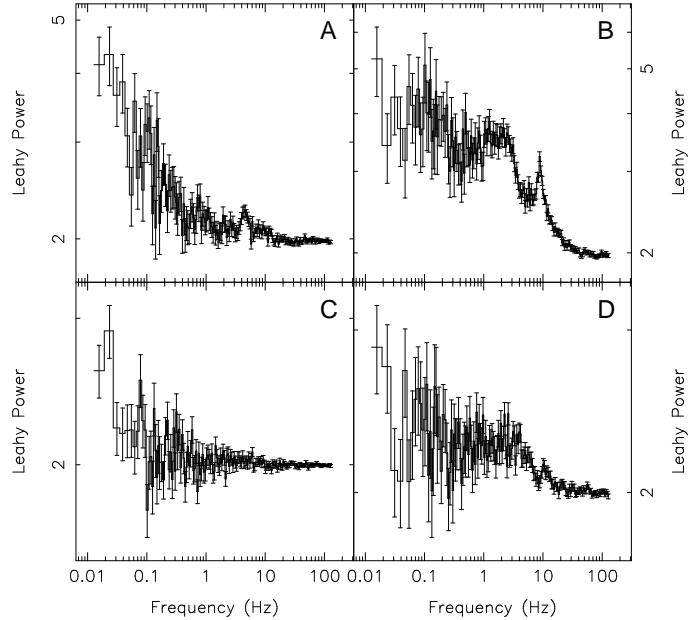}}
\caption{\scshape \small Four power spectra during the decay: (a) bottom of branch III
(MJDs 51269 and 51273), (b) top of branch III (MJDs 51270 and 51271),
(c) bottom of branch IV, and (d) top of branch IV. Poisson level was
not subtracted.\label{decay-pds_fig}}
\end{figure}

\begin{figure}[t]
\centerline{\psfig{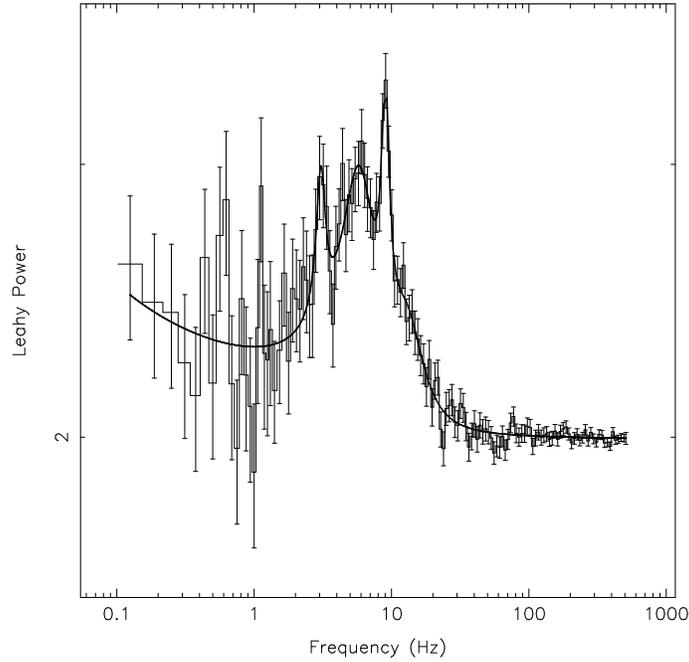}}
\caption{\scshape \small The 6.5--60 keV power spectrum of MJD 51271, showing similar
structure as found in the type B power spectra. The solid line shows
the best fit with four QPOs and a power law. Poisson level was not
subtracted.\label{obs_72_fig}}
\end{figure}

The next two observations (MJDs 51270 and 51271) were located near the
top of branch III. Their power spectra were very similar. The MJD
51270 power spectrum showed a QPO at 8.9$\pm$0.1 Hz
(1.8$^{+0.2}_{-0.1}$\% rms, FWHM=2.1$^{+0.5}_{-0.3}$ Hz) and a broad
peak around 2 Hz that was fitted with a Gaussian at 1.8$\pm$0.1 Hz
(3.2$\pm$0.2\% rms, FWHM=2.9$\pm$0.4 Hz) plus an exponentially cutoff
power law (4.4$\pm$0.3\% rms, $\alpha=-$1.6$\pm$0.6,
$\nu_{cutoff}=4\pm$1 Hz). The power spectrum of MJD 51271 showed a QPO
at 9.05$\pm$0.12 Hz (1.8$^{+0.3}_{-0.2}$\% rms,
FWHM=1.4$^{+0.5}_{-0.4}$ Hz) and a broad peak around 2 Hz that was
fitted with a Gaussian at 1.7$\pm$0.3 Hz (3.8$^{+0.4}_{-0.3}$\% rms,
FWHM=3.74$\pm$0.5 Hz) plus an exponentially cutoff power law
(4.5$\pm$0.4\% rms, $\alpha=-$2.3$\pm$1.0, $\nu_{cutoff}=3\pm$1
Hz). The combined 1/128--128 Hz power spectrum of the two observations
is shown in Figure \ref{decay-pds_fig}b. The strength of the 0.01--1
Hz noise, which was fitted with a power law, was $\sim$1.5\% rms, but
it should be noted that some of the power in the 0.01--1 Hz range was
absorbed by the Gaussian and the exponentially cutoff power law. The
two $\sim$9 Hz QPOs had relatively high Q-values (4.2 and 6.5), which
suggests that they were of type B; this seems to be confirmed by the
shape of the power spectra at higher energies; Figure \ref{obs_72_fig}
shows the 6.5--60 keV power spectrum of MJD 51271, which could be
fitted with a power law and QPOs at 3.1$\pm$0.1 Hz, 5.7$\pm$0.2 Hz,
9.0$\pm$0.1 Hz, and 12.5$\pm$1.0 Hz. This is reminiscent of the type B
QPO found on branch II, and the IS power spectrum shown in Figure
\ref{obs1_fit_fig}.  In the combined 2--60 keV power spectrum of the
two observations a QPO at 251$\pm$3 Hz was found. It had an rms
amplitude of 2.21$\pm$0.15\% and a FWHM of 42$\pm$6 Hz. Its location
in Figure \ref{low-high-qpo_fig} is shown by a diamond. Although the
frequency of the QPO lies in the $\nu_{HF}$ range found on branch II,
the count rate at which is was found was considerably lower
($\sim$1350 counts s$^{-1}$ compared to 4700--8300 counts s$^{-1}$ on
branch II).

The power spectrum of the next observation (MJD 51273), which in the
CD was located close to the MJD 51269 observation, showed a QPO at
4.52$\pm$0.13 with an rms amplitude of 1.2$\pm$0.02\% and a FWHM of
1.1$^{+0.5}_{-0.3}$ Hz. The QPO is most likely of type A, based on its
strength and lack of harmonic structure. The low frequency noise had a
strength of 1.20$\pm$0.05\% rms. The combined power spectrum of MJDs
51269 and 51273 is shown in Figure \ref{decay-pds_fig}a.

The observation on MJD 51274 showed no QPOs, and the 0.01--1 Hz noise
had an rms amplitude of less than 0.4\%. During MJD 51274--51280
(HC$\sim$0.007, SC$\sim$0.13) the source showed similar power spectra
that, when combined, were fitted with a single power law with a
strength 0.90$\pm$0.09\% rms and an index of 1.0$\pm$0.2, which is
typical for the HS.

Between MJD 51283 and MJD 51298 XTE J1550--564 traced out branch IV in
the CD.  At the top of the branch (SC$>$0.35) the power spectra showed
a QPO and a peaked noise component. These were not found at the bottom
of the branch (SC$<$0.35). The combined power spectrum of the bottom
of branch IV (Fig.  \ref{decay-pds_fig}c) was fitted with a power law
with an rms amplitude of 2.4$\pm$0.2\% and an index of 0.7$\pm$0.1. In
the combined power spectrum of the top of branch IV
(Fig. \ref{decay-pds_fig}d) a QPO was found at 11.3$\pm$0.5 Hz, with
an rms amplitude of 4.1$^{+0.7}_{-0.5}$\% and a FWHM of
2.9$^{+1.1}_{-0.8}$ Hz. A peaked noise component was present below 10
Hz. It was fitted by a Lorentzian with a frequency of 3.0$\pm$0.3 Hz,
an rms amplitude of 10.1$^{+0.9}_{-0.8}$\%, and a FWHM of 5.7$\pm$0.9
Hz. The 0.01--1 Hz noise was fitted with a power law that had a
strength of 2.8$\pm$0.4\% and an index of 0.3$\pm$0.1.

Both branch III and IV showed behavior that was similar to that seen
in the VHS (branch II). Since the count rates were lower than on the
branch II, these branches were probably IS (at least, when QPOs were
seen).

\begin{figure}[t]
\centerline{\scshape \small \psfig{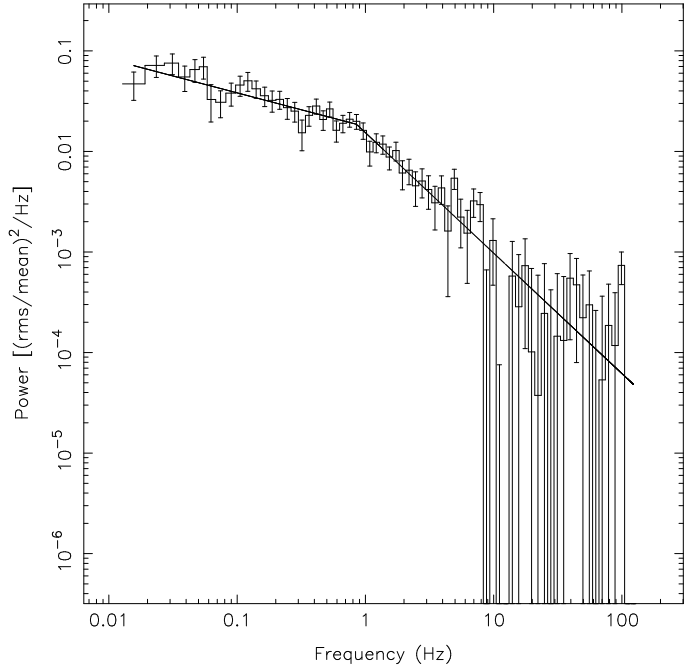}}
\caption{\scshape \small Combined power spectrum of the last six observations during
the low state (MJD 51307--51318). The solid line is the best fit with
a broken power law. The Poisson level was
subtracted.\label{ls-pds.ps}}
\end{figure}

Between MJD 51299 and MJD 51318 branch V was traced out in the CD. It
reached much harder colors than before, and the movement up the branch
was accompanied by a considerable increase in the strength of the low
frequency noise, as can be seen from Figure
\ref{noise_curve_fig}. There was a clear difference between the power
spectra at the top and bottom of the branch.  The combined MJD
51299--51306 (bottom part of branch V, SC$<$0.6) power spectrum was
fitted with a power law with an rms amplitude of 8.9$\pm$1.0\% and an
index of 0.7$\pm$0.1. A single power law yielded a poor $\chi^2_{red}$
(3.2 for d.o.f.=67) for the combined MJD 51307--51318 (top of branch
V, SC$>$0.6) power spectrum, and a broken power law was used instead
(see Figure \ref{ls-pds.ps}). Its rms amplitude was 15.9$\pm$0.3\%,
with $\nu_{break}=0.9\pm0.1$ Hz, $\alpha_1=0.33\pm0.04$, and
$\alpha_2=1.2\pm0.1$ ($\chi^2_{red}=1.1$ for d.o.f.=67). The strength
and shape of the noise, and the spectral hardness suggests that the
bottom of branch V the source was in the IS, and that at the top of
branch V the source was in the LS.

\section{Radio Observation}\label{radio_sec}

On 1999 March 11 (MJD 51248) we observed the radio counterpart of XTE
J1550--564 \citep{camchu1998} with the Australia Telescope compact
array (ATCA), in a high-resolution 6 km configuration.  Observations
were made simultaneously at 6.3 and 3.5 cm, and at 21.7 and 12.7 cm,
in order to obtain broad band spectral coverage. The observations were
interleaved with those of a nearby reference source B1554-64, for
phase calibration every 25 min. The source was clearly detected at all
four wavelengths; the mean flux densities at 21.7, 12.7, 6.3 \& 3.5 cm
were, respectively, 5.1, 3.0, 2.8, and 1.9 mJy (errors
$\sim$10\%). The four flux densities were fitted with a power law
corresponding to a spectral index ($\alpha=\Delta log S_\nu/\Delta log
\nu$) of  $-$0.53$\pm$0.12.  The location of the MJD 51248 RXTE
observation is indicated with by `ATCA' in Figure \ref{cd_fig}.

\section{Discussion}\label{discuss_sec}

In this section we present a discussion of our results. We start by
briefly summarizing the results. After that the source states and
power spectra are discussed.

\subsection{Summary of Results}\label{sum_sec}

In the period of 1998 November 22 to 1999 May 20 XTE J1550--564 showed
a wide variety of behavior. To organize the different phenomena and
relate them to each other, it is useful to compare the rapid time
variability and the energy spectra. An initial division of the
observations can be made based on the strength of the broad band
(1/128--128 Hz) power, which is shown in Figure
\ref{noise_curve_fig}b. It can be seen that the source alternated
between states with low power (a few percent rms) and states with high
power (more than a few percent). When comparing this figure with
Figure \ref{lc_fig} it is obvious that observations with high power
were mainly found when the spectrum was hard. These spectrally hard
states appeared as branches (I--V) in the color-color
(Fig. \ref{cd_fig}) and hardness-intensity (Fig. \ref{hid_fig})
diagrams. From the hardness-intensity diagram it is apparent that the
hard branches occurred at five distinct count rate levels, and that
they were separated by periods that were spectrally soft(er); in the
color-color diagram the hard branches lay more or less parallel to the
power law curve. The power spectra on the five hard branches often
showed QPOs, and in some cases also strong peaked and/or band limited
noise (Fig. \ref{qpo-types_fig}).  These properties classify the
observations on the branches as VHS, IS, or LS. The power spectra that
were not on the hard branches showed noise with strength of a few
percent rms and, when combined, a weak QPO around 17 Hz
(Fig. \ref{fl-int_pow_fig}). These observations can be classified as
HS; in the color-color diagram they lay close to the disk blackbody
curve. When the source moved up a hard branch the low frequency noise
changed from a weak power law to strong band-limited. On branch II
this change was accompanied by the QPOs changing from the broad type
A, to the narrow types B and C (Fig. \ref{qpo-hc_fig}).  On branches
II and III we also found high frequency QPOs in the 102--284 Hz
range. Their frequencies were anti-correlated with the hardness of the
energy spectrum (Fig. \ref{qpo-hc_fig}), and correlated with the
frequency of the type A, B, and C low frequency QPOs
(Fig. \ref{low-high-qpo_fig}).

\subsection{Source States}\label{states_sec}

In recent years the picture of black hole behavior that emerged from
observations was consistent with a one-dimensional scheme, in which
four canonical source states were linked by one parameter, usually
taken to be the mass accretion rate (see, e.g. \citet{esmcna1997} and
\citet{esnacu1998} for recent elaborations on this view). In the
context of two-component spectral models, often interpreted in terms
of emission from an accretion disk and a hot comptonizing medium, this
implies that both components contribute to the energy spectrum and
power spectrum in amounts that depend strictly on this parameter; if
both components were to vary independently, the description of the
phenomenology would have to be at least two-dimensional.  XTE
J1550--564 seems to provide evidence for such two-dimensional
behavior. Although the source was observed in all four canonical
states, their occurrence was more complex than expected on the basis
of a simple relation with the mass accretion rate.

We start by discussing the relation between the source states and the
position of the source in the CD and HID (see Figures \ref{cd_fig} and
\ref{hid_fig}).  The motion of the source through the CD was along
branches.  One branch (hereafter the soft branch) lay parallel to the DBB
curve in the CD, and quite close to it (Fig \ref{cd_fig}a). Whenever
XTE J1550--564 was on or close to this branch (e.g. flares 1-5), it
could be classified as being in the HS: it showed soft energy spectra,
and the (power law-like) low frequency noise had a strength of only a
few percent rms. All the other branches (hereafter hard branches) lay
approximately parallel to the power law curve in the CD. In time,
these hard branches were traced out one after the other, and all but
the last two were clearly separated from each other by intervals that
showed HS behavior. When the source was on a hard branch it was in the
VHS, IS or LS: the energy spectrum was hard, and the power spectrum
showed QPOs and/or strong noise.  The hard branches were similar to
each other in that the shape of the noise changed from power law like
to band limited as the source moved up such a branch (i.e. when it became
harder). When the source was on a hard branch still relatively close
to the soft branch it would, based on the energy and power spectrum,
usually be classified as being in the canonical HS - only further up the
hard branches full-fledged VHS, IS and LS behaviour emerged.
Canonical LS (variability) behavior was only found at the top of
branch V, the branch that reached the hardest colors.

Although the observations on hard branches I, III, and IV were
classified as IS, and those on hard branch II as VHS, they had in
fact very similar properties, the only difference being the count rate
at which they were observed. This was already found for the IS and VHS
in other sources, e.g., GS 1124--68 and GX 339-4
\citep{bevale1997,meva1997}. We therefore regard the VHS as an
instance of the IS, but just the one that happens to be the brightest.

The behavior on the hard branches that were traced out before our
observations, during the first part of the outburst, was similar to
that during our observations \citep{remcso1999,cuzhch1999}. During the
first part of the outburst LS behavior (strong band limited noise with
a break around 0.1 Hz) was observed only when the hardness was similar
to that at the top of branch V (at the start of the outburst, when the
count rate was $\sim$100 times as high as on branch V). Moreover, the
source evolved from LS to the HS via a VHS (or IS as we shall
henceforth call it), clearly showing the same ordering of states (as a
function hardness) as during our observations.

The above shows that as the hardness increased the source evolved from
the HS via the IS, to the LS; the hard branches therefore corresponded
to HS$\leftrightarrow$IS$\leftrightarrow$LS transitions, or at least
attempts to, since not every branch reached the LS and the source did
not always return completely to the HS. Similar conclusions were also
drawn by \citet{ruleva1999}, on the basis of a comparative study of 10
black hole candidates. They found that the VHS and IS were spectrally
intermediate to the HS and LS, and also that as the hardness increases
the noise switches from HS-like to LS-like. They also concluded that
transitions between the HS and LS could take place at
luminosities/count rates both above and below that of the HS. In our
observations transitions between HS and IS were found at around 8000,
1200, 600, and 200 counts s$^{-1}$ (see Fig. \ref{hid_fig});
transitions between IS and LS were found at around 40
(Fig. \ref{hid_fig}) and 4000 counts s$^{-1}$ (during the first part
of the outburst). Moreover we observed HS behavior at all count rate
between 200 and 10000 count s$^{-1}$. All this argues against a
one-dimensional description of the state transitions as a function of
the mass accretion rate.

The observations of XTE J1550--564 contradict the old picture of black
hole states, in which hard states are only found at the highest and
lowest count rates. XTE J1550--564 clearly shows that hard states can
be observed at any count rate level. However, some remarks should be
made. All the hard states (I--V) were observed during the decay of the
source (I during the decay of the first part of the outburst, II--V
during the decay of the second part). Also, the intervals between
branches II--IV, although they could be classified as HS, had
significantly harder spectra than the HS observations during the rise,
which were extremely soft. This suggests that the conditions for the
presence of the hard spectral component are more favorable during the
decay or phases of low count rate. This is supported by the fact that
all small flares (1--5) were observed at count rates higher than that
of the hard branches. The fact that these flares did not develop into
real transitions suggests that the conditions for transitions and
development of the hard spectral component are less favorable at the
highest count rates. Also, it can be argued that the only LS that was
observed during our observations was found at the lowest count rates
at the end of the outburst, as expected in the canonical picture of
black hole states. On the other hand, a LS was also found during the
first part of the outburst when the count rate was at least a factor
100 higher than during the LS at the end of the outburst, clearly
showing that LS is not only found at the lowest count rates.

\begin{figure}[t]
\centerline{\psfig{figure=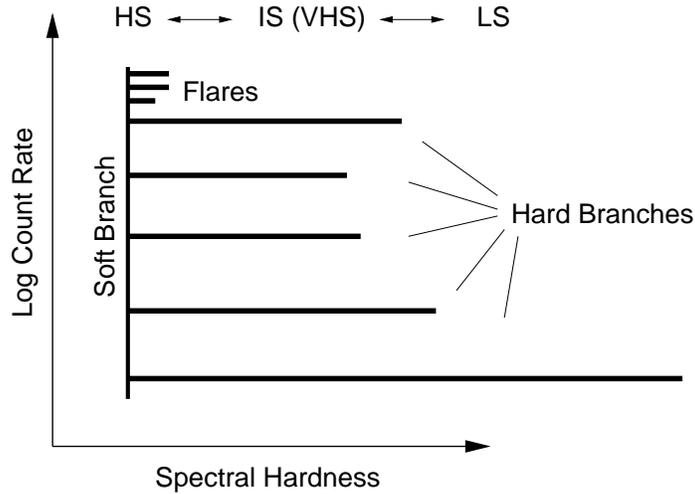,width=9cm}}
\caption{\scshape \small A diagram of the two-dimensional behavior of
XTE J1550--564 during the second part of its 1998/1999 outburst. In
the color-color and hardness-intensity rate diagrams the source traced
out a comb-like structure, with the spectrally soft state (HS) being
the regions on and near the spine, and the
HS$\leftrightarrow$IS(VHS)$\leftrightarrow$LS transitions being the
teeth. The comb-like structure in the CD and HID is caused by three
effects. First, the disk and power law luminosity change to a large
extent independently. Second, the power law luminosity can be
negligible for times comparable to the time scale on which the disk
luminosity changes. Third, the time scale on which the power law
luminosity changes, when it is not negligible, is short compared to
that of the disk luminosity changes. As the relative contribution of
the power law component increases, the low frequency noise changes
from HS-like (weak power law) to LS-like (strong band limited).  In
this scheme it is possible to have a soft state (HS) at a higher disk
luminosity than a the brightest IS (VHS). Although the structure
traced out by XTE J1550--564 was comb-like, similar two-dimensional
behavior of other black holes may lead to different
structures.\label{diagram_fig}}
\end{figure}

Based on the CD, HID, and power spectral fits one gets the impression
that the behavior of XTE J1550--564 is two-dimensional, i.e. at least
two (observable) parameters are needed two describe the appearance of
the source and to account for the occurrence of the different states.
Phenomenologically, the two parameters describing this two-dimensional
behavior are the count rate and the spectral hardness.  A schematic
representation of the behavior of XTE J1550--564 in terms of these
parameters is shown in Figure \ref{diagram_fig}. It shows that the
states are arranged in a comb-like topology, with the soft (HS) branch
being the spine, and the HS$\leftrightarrow$IS$\leftrightarrow$LS
transitions being represented by the teeth. Note that the parameter on
the vertical axis is the logarithm of the count rate, and that we
decided to show the hard branches as horizontal lines; for reasons of
clarity we did not depict the exact movement of the source through the
diagram.  We also included the location of the flares, which occurred
at count rates higher than that of the brightest hard branch.

The most important aspect of XTE J1550--564 is probably the fact that
two observable parameters (count rate and hardness) varied too a large
extent independently from each other.  This suggests that at least two
physical parameters underlie this behavior, since this complex
behavior is hard to explain within a framework where the appearance of
the source is determined by a only single physical parameter
(e.g. only mass accretion rate). Physically the two underlying
parameters might for example be the mass accretion rate through the
disk (roughly increasing with count rate, at least in the HS) and the
size of a Comptonizing medium (increasing with spectral hardness). In
that case the state of the source is determined by the (relative) size
of the Comptonizing medium, with it being small or absent in the HS
and growing in size towards the LS. The fact that we see increases in
the spectral hardness at many count rate levels suggests that the size
of this medium, and therefore also the state of the source, is to a
large degree not determined by the accretion rate through the disk.
We note that the inner disc radius as derived from variability
properties (i.e. QPO frequencies, see Section \ref{disc_low_sec})
correlates well with hardness, suggesting that the Comptonizing medium
grows as the inner disk edge moves out. The two physical
parameters do not necessarily vary completely independently from each
other.  For instance, changes in one parameter may be triggered by
changes in the other, and certain values of one parameter may restrict
the value of the other parameter (e.g. in between the hard branches
XTE J1550--564 seemed to become slightly harder towards lower count
rates (Fig.  \ref{hid_fig})).

While previous authors, inspired by the description of black hole
spectra in terms of two components, have also discussed black hole
phenomenology in terms of two-dimensional diagrams
\citep{mikiig1994,no1995}, the overall picture has in our view been
considerably clarified by the clues provided by XTE J1550--564
described in this paper. The basic phenomenology seems to be one where
the hard and soft component can vary to a large extent independently
from each other. The HS is the name we have given in the past to all
cases where the hard component is weak compared to the soft component,
and the IS/VHS and LS are unified into cases where the hard component
is, respectively, comparable to or dominating the soft component.  The
difference between the VHS and IS is reduced to a difference in the
luminosity of the soft component at which they occur, and the
difference between IS/VHS and LS is caused by differences in the
relative contributions of the soft and hard components.

This two-dimensional interpretation might very well be applicable to
all black hole candidates showing the canonical states. The often
observed order of states
(VHS$\rightarrow$HS$\rightarrow$IS$\rightarrow$LS) is still consistent
with the diagram drawn Figure \ref{diagram_fig}. The fact that black
hole state behavior often seems one-dimensional can be explained by
considering how the behavior of XTE J1550--564 would have appeared if
the data were of lower quality, the source sampling was more
infrequent, its distance was larger, and if it had different
characteristic time scales for variations in the soft and hard
component. Much of its subtle behavior would have been missed or may
have been misinterpreted. It is mainly thanks to the combination of
source brightness, the quality of the RXTE/PCA data, and the excellent
source sampling that we clearly see the two-dimensional nature of its
behavior. Since the observations of XTE J1550--564 strongly suggest
that mass accretion rate through the disk and state are decoupled, it
would be possible to see transients that remain in the same state
during a whole outburst. Moreover, one could also see state
transitions in sources with a more or less constant mass accretion
rate. Suggestions of such behavior have been seen in, respectively, GS
2023+338 \citep{sukaef1991,temiki1992,mikiig1992} and Cyg X-1
(\citet{zhcuha1997}, however see \citet{frpazd2000}).

The radio brightness of XTE J1550--564 during the MJD 51248 ATCA
observations indicates that an outflow event was going on or had
recently occurred. The spectral index suggests that the radio source
was optically thin, and observed during the decay of such an outflow
event.  This event might be associated with the state change on MJD
51237/51239 (the onset of branch II).  Although XTE J1550--564 was
observed in radio only once (on branch II), observations of other
black hole candidates \citep{fecotz1999,fe2000} suggest that radio
emission is associated with spectrally hard states. The hard branches
may therefore correspond to changes in the accretion flow geometry,
where an inflow (HS) is gradually accompanied by (or changing into) an
outflow (LS). Jet-like outflow models have already been proposed for
the VHS in GX 339--4 by \citet{miki1991}.

\subsection{Power spectra}\label{pow-spec_sec}

Although not always observable in individual observations, QPOs were
found on all branches, except for the last one. Several types were
found: 1--18 Hz QPOs on the hard branches (type A, B, and C), 15--18
Hz (plus an harmonic) on the soft/HS branch and in the flares, and
100--285 Hz QPOs on the two brightest hard branches.

\subsubsection{High frequency QPOs}

High frequency QPOs in black hole candidates are a relatively new
phenomenon. Previous to XTE J1550--564, they were found in GRS
1915+105 \citep[67 Hz]{moregr1997} and in GRO J1655-40 \citep[300
Hz]{remomc1999}. \citet{remcso1999} found high frequency QPOs in XTE
J1550--564 in the 161--238 Hz range, during the first part of the
outburst. With the observations of the second part of the outburst
this range has been expanded to 100--285 Hz.  It is obvious that the
frequency of the high frequency QPOs in XTE J1550--564 can not be
explained by models that predict an approximately constant frequency,
e.g. orbital motion at the innermost stable orbit \citep{moregr1997},
Lense-Thirring precession at the innermost stable orbit
\citep{cuzhch1998}, or trapped-mode disk oscillations
\citep{nowabe1997}. 

The high frequency QPO was found on two branches; between 100 and 285
Hz on branch II, and at 251 Hz on branch III. The count rates at which
it was observed, were much lower on branch III ($\sim$1350 counts
s$^{-1}$) than on branch II ($\sim$4700-8300 counts s$^{-1}$), which
shows that the QPO frequency does not strongly depend on the count
rate. A similar effect is also seen for the high frequency QPOs in
some neutron star X-ray binaries \citep{mevafo1999,fovame2000}, where
the frequency varies along parallel branches in a frequency--count
rate diagram. A certain frequency can therefore be observed at
different count rate levels, and a range of frequencies can be found
within a relatively small range of count rates. This suggests that if
the QPO frequency is related to a certain (variable) radius, this
radius varies almost independently from the count rate (and probably
mass accretion rate; \citet{va2000}). Similar two-dimensional behavior
as discussed in Section \ref{states_sec} may therefore also be present
in some of the neutron star sources.

An obvious question to ask is, whether the high frequency QPOs in
black hole candidates have the same origin as the kiloHertz QPOs that
are observed in the neutron star sources (see \citet{va2000} for a
review). Of course, since the QPOs in the neutron star sources are
often observed in pairs, only one (if any) of those two QPOs can have
the same origin as the QPOs in black hole sources, which until now
have always appeared as single peaks. It is, however, not clear which
of the two QPOs that would be; both the lower and upper kHz QPO have
Q-values and the rms energy spectra that are consistent with those of
the QPO in XTE J1550--564.  The frequency ranges in which the high
frequency QPOs are observed are 102--284 Hz for the QPO in XTE
J1550-564, and 200--1070 Hz for the lower kHz QPO, and 325--1330 for
the upper kHz QPO in the neutron stars. Here we combined the kHz QPO
data for all neutron star sources in \citet{va2000}. Although the
lower and upper kHz QPOs cover a frequency range of a factor of 5.4
and 4.1, respectively, the values for individual neutron star sources
are more like that found for XTE J1550--564 ($\sim$2.8).

Although the high frequency QPO in XTE J1550--564 has parameters that
are consistent with those of both the lower and upper kHz QPO (within
a simple orbital frequency model), a major difference between XTE
J1550--564 and the neutron star sources is the fact that the latter
often show two high frequency QPOs. However, this could be 
explained if  one of the two kHz QPOs in the neutron star
sources is due to a mechanism that requires the presence of a solid
surface.

\subsubsection{Low frequency QPOs}\label{disc_low_sec}

On all the hard branches, except branch V, QPOs were found with
frequencies between 1 and 18 Hz. Due to the low count rates the
quality of the power spectra on branch V was poor, and the presence of
QPOs could therefore not be ruled out. The frequencies of the QPOs
were not constant, as can be seen from Figure \ref{qpo-hc_fig}
(showing the QPOs found on branch II), but it is not immediately clear
from that figure how the frequency of the low frequency QPOs
($\nu_{LF}$) depended on the hard color (which is a good measure of
the distance along the branch). Although the figure is suggestive of a
positive correlation, especially for HC$<$0.015, the behavior of the
low frequency QPOs during the rise of the first part of the outburst
of XTE J1550--564 (anti-correlation with hardness; \citet{cuzhch1999})
and the correlation found between the low and high frequency QPOs (see
below, and Figure \ref{low-high-qpo_fig}) lead us to believe that
hard color and frequency were anti-correlated, and that the
frequencies of both the high and low frequency QPOs decreased as the
source moved up branch II. This contradicts the switch from a
correlation into anti-correlation when the hard color passes a certain
value, that was reported by \citet{ruleva1999} for other black hole
candidates. We want to stress again that the fact that several
harmonics were present and that not always the same harmonic was the
strongest one, can easily lead to confusion. On the other hard
branches not enough QPOs were observed to confirm the anti-correlation
with hardness.

It was usually the QPO that happened to be located between 5 and 8 Hz
that was the strongest in the 2--60 keV power spectra of branch II,
even though it could be identified with different harmonic components
(as can be seen from Figures \ref{qpo-hc_fig} and
\ref{low-high-qpo_fig}). Perhaps variations within that frequency
range are less prone to damping than outside, or a resonance
occurs. The shoulder component that was present at a frequency of 1.25
times that of the 5--8 Hz type B QPOs, has been found before in QPOs
in other black hole candidates like GS 1124--68 and GX 339--4
\citep{bevale1997}, but recently also in the 20-50 Hz QPOs in the
neutron star system GX 340+0 \citep{jovawi2000}.

Figures \ref{qpo-hc_fig} and \ref{low-high-qpo_fig} seem to indicate
that the low frequency QPOs evolve from type A via type B into type C,
and vice versa. On branch II the type A QPOs were located at the
bottom of the branch, and the those of type B and C further
along it. This was also seen on branch III for type A and B
(Figs. \ref{decay-pds_fig} and \ref{obs_72_fig}), and to certain
extent also on branch I, where indications for type B QPOs were found
at a similar hardness as where they were found on branch II
(Fig. \ref{obs1_fit_fig}). No direct transitions between type A and B
were seen, so it is not known whether such transitions are smooth or
abrupt. In the MJD 51254 observation a jump in the count rate was
accompanied by a change in the power spectrum that may have been a
transition from type C to B. This suggests that the transitions between
the different types are quite sudden. Apart from the difference in
spectral hardness at which type A and B occurred, it is clear that
there are at least two other fundamental differences between the two
types. First there is the difference in the Q-value, which is higher
in type B. This might also explain why more harmonics are detected in
the type B power spectra, since narrower features are easier to
identify. Second, there is the difference in time lag spectra
\citep{wihova1999}, which can not be reconciled even when one compares
the time lags of similar harmonics (i.e., the type A-I 12 Hz QPO and
the type B 6 Hz QPO, which in our view are supposed to be the same
harmonic, still have the opposite signs for their time lags).  As
mentioned already in Section \ref{vhs} the type A QPOs were divided in
two types, A-I and A-II. Figure \ref{qpo-hc_fig} shows that type A-II
QPOs (triangles) were found both at higher and lower hardness than
type A-I (circles). The A-II subtype should therefore be regarded as a
collective rather than a real type. The two filled triangles
(HC$\sim$0.012) can be identified with second harmonics and are for
that reason different from the A-I observations, where the fundamental
was the dominant harmonic. The nature of the two observations
represented by the open triangles in Figure \ref{qpo-hc_fig} remains
uncertain; their frequency is lower than expected on the basis of
their hardness and in one case an addiational broad bump was observed
in the power spectrum.

We found that the frequencies of the high frequency QPOs and low
frequency QPOs that were detected simultaneously are well correlated
(Fig. \ref{low-high-qpo_fig}); it is mainly on the basis of this that
we conclude that the different types (A, B, and C) of low frequency
QPOs have the same origin. Similar correlations have also been found
for the low and high frequency QPOs in a number of neutron star source
(e.g. \citet{vaswzh1996,vawiho1997,wihova1997,jowiva1998,fova1998,pswiho1999,mastsw1999})
and in neutron star and black hole sources for QPOs and broad noise
components \citep{psbeva1999}. The main correlation that was found for
the neutron star and black hole sources extended over a frequency
range of 0.1--1200 Hz, and was consistent with the relation
$\nu_{LF}=(42\pm3)(\nu_{HF}/500\, Hz)^{0.95\pm0.16}$, that was
found in the neutron star Z sources \citep{pswiho1999,psbeva1999}. A
second correlation was present in Figure 2 of \citet{psbeva1999} that
was fitted with $\nu_{LF}=2.09\times10^{-3}(\nu_{HF})^{1.46}$ by
\citet{di2000a} for data of 4U 1728--34. Both relations are plotted in
Figure \ref{low-high-qpo_fig} (dashed and dotted lines, respectively),
and are apparently not consistent with the data of XTE J1550--564.

The fact that the data in Figure \ref{low-high-qpo_fig} are well
fitted with four linear relations that do not pass through the origin
excludes models in which $\nu_{LF}$ and $\nu_{HF}$ are related by a
simple power law expression. The four linear fits to the data in
Figure \ref{low-high-qpo_fig} cross each other at $\nu_{LF}\sim$0 Hz
and $\nu_{HF}\sim$ 75 Hz.  It is not clear what the nature of this
$\sim$75 Hz frequency in a black hole system could be.

Low frequency QPOs were also found on the soft branch (15.6 Hz) and in
the high state flares (17.9 Hz). They were much weaker than the low
frequency QPOs found on the hard branches and had a rather high
Q-value ($\sim$10). Their frequency was apparently correlated with
hard color (assuming that we observed the same harmonical component),
unlike that of the hard branch QPOs. The above suggests that the 15.6
Hz and 17.9 Hz QPOs may have a different origin than the type A, B and
C QPOs, and it may also explain why the only 18 Hz QPO (MJD 51239)
reported by \citet{somcre2000a} did not follow the relation between
QPO properties and the spectral parameters seen on the hard
branches. On the other hand, there some clues that do suggest a
relation with the type A, B, and C QPOs: when their values would be
plotted in Figure \ref{qpo-hc_fig} they would lie close to the
extrapolation of a line through the filled symbols. Moreover, the 15.6
Hz and the 17.9 Hz QPO both fall on the empirical relation found by
\citet{wiva1998} for the low frequency QPO and break frequency found
in many types of X-ray binaries, including black holes. Hence, it is
not clear whether these QPO really have a different origin than the A,
B, and C type QPOs, or that they only have appear to be different
because the hard spectral component is so much weaker.  The properties
of the 16--18 Hz QPO are in fact remarkably similar to those of the
14--23 Hz QPO in GRO J1655--564, studied by \citet{somcre2000a}; the
similarity extends to the frequency at which they are found, their
amplitude, and their relation with the disk spectral parameters.  Note
the apparent switch from a correlation of QPO frequency with hardness
(15.6 Hz and 17.9 Hz QPOs) to an anticorrelation (A, B and C types) is
probably not related the one reported by \citet[see
above]{ruleva1999}. They did not find QPOs in such spectrally soft
states, and the frequencies of the 15.6/17.9 Hz QPOs are still well
above that of most A, B, and C type QPOs.

The range over which the fundamental of the low frequency QPOs is
observed in XTE J1550--564 is 0.1 Hz to 6 Hz (which includes the QPOs
reported by \citet{cuzhch1998}, but not the 16--18 Hz QPOs discussed
in the previous paragraph). Using the expression for the lowest line
in Figure \ref{low-high-qpo_fig}, $\nu_{HF} = 38.1(\nu_{LF} + 1.61)$,
and assuming that the high frequency QPO is due to orbital motion at
inner disc radius ($R_{in}$), $\nu_{HF}\propto R_{in}^{2/3}$, one can
estimate the corresponding changes in the inner disc radius. For the
low frequency QPO changing from 0.1 to 6 Hz we find a decrease in
$R_{in}$ by a factor 2.6, which is comparable to what was found by
\citet{dips1999} for other black hole systems. However, their relation
between $R_{in}$ and $\nu_{LF}$ was based on the empirical relation
between $\nu_{LF}$ and $\nu_{HF}$ found for neutron star sources by
\citet{psbeva1999}.  Using the relation of \citet{dips1999} we find a
decrease in $R_{in}$ by factor $\sim$4.2. Both numbers suggest that
the inner radius changes are rather small when a black hole changes
from a hard (where the 0.1 Hz QPO was observed) to a much softer state
(where the 6 Hz QPO was observed). It should be noted that these
changes may in fact be somewhat larger if the 15.6/17.9 Hz QPO turns
out to be related to the A, B, and C type QPOs, and/or if the lowest
peak in the \citet{cuzhch1998} power spectra is not the fundamental,
but a higher harmonic.

There are two types of low frequency QPOs in the neutron star Z
sources \citep{hava1989} that may be compared with the low frequency
QPOs in XTE J1550--564 (and other black hole candidates): these are the
horizontal branch QPOs (HBO) and the normal branch QPOs (NBO) (see
\citet{va1995b} for a review). Similar QPOs have also been found in a
number of neutron star atoll sources. Of the two QPO types in Z
sources it is the HBO that bears most resemblance to the QPOs in XTE
J1550--564. Unlike NBOs, HBOs  have a strong harmonic content;
e.g., in GX 340+0 the HBOs could be fitted with three harmonically
related peaks (1st, 2nd and 4th) plus a shoulder component for the
second harmonic \citep{jovawi2000}, similar to the type B QPOs in XTE
J1550--564. HBOs are found in the 15--60 Hz range, and their frequency
changes smoothly; NBOs are found in the 6--20 Hz range, but their
frequency changes are strongly related to sudden spectral/state
changes. Although the 6--20 Hz range of the NBOs is closer to the
values we found for the low frequency QPOs in XTE J1550--564, it
should be noted that if one scales the frequency of those low
frequency QPOs with a factor $\sim$5 (which is maximum $\nu_{upper}$
in the neutron star sources divided by the maximum $\nu_{HF}$ in XTE
J1550--564), one gets values that are similar to those found for the
HBO. Another indication that the QPOs in XTE J1550--564 are related to
the HBOs (at least to the HBOs in the Z source GX 17+2
\citep{di2000b,ho2001} ) is the fact that they are both observed when
a hard power law tail is present in the energy spectrum, and that both
show an anti-correlation between their frequency and the strength of
this high energy component. Finally we note that the HBO in Z sources
is accompanied by a low frequency noise component that becomes
stronger when the HBO frequency decreases, which is similar to what is
observed in XTE J1550--564, where a strong noise component develops
when the QPO frequency drops (see also figures in \citet{cuzhch1999}
and \citet{remcso1999}).

When comparing the properties of the 0.01--0.1 Hz noise in the high
state flares with those in between the flares, we find that in the
flares the noise was stronger, but had a softer fractional rms
spectrum, whereas the overall X-ray spectrum of the source was
harder. Though this might at first appear remarkable, it is in perfect
agreement with the assumption that the extra noise is associated with
the hard power law component. A hard spectral component, with
associated noise that remains a constant fraction of it, in
combination with a soft spectral component that remains unchanged,
will lead to a softer fractional spectral dependence of the noise when
the hard component increases.

The transition observed on MJD 51254 showed many similarities to the
``dips'' and ``flip-flops'' observed in GX 339--4 and GS 1124--68
\citep{miki1991,tadomi1997}, although the time scale of the transition
we observed ($\sim$100 s) is quite long compared to the transitions in
these dips and flip-flops. Both showed a QPO in their upper count rate
level (but not in their lower level), and a somewhat stronger noise in
their lower count rate level.  Like in GX 339--4 and GS 1124--68 the
transition occurred in a region in the CD where power law noise
changes to band limited noise. Also, the count rate differences
($\sim$10\%) were accompanied by relatively subtle spectral
differences, showing that on these short time scales spectral hardness
and power spectral properties do not correlate as well as they do on
longer time scales. The transition probably originated in the
accretion disc component; the frequency of the QPOs before and after
the transition were different and indicate that the inner disk radius
had decreased a few percent. This change did apparently not affect the
spectrum of the disk component, since the soft color remained
constant. The hard color on the other hand did change, but certainly
not as dramatically as the count rate. A slow decrease in the hard
color started around the time of the transition maybe reflecting some
kind of cooling of the hard spectral component.

\section{Conclusions}\label{conlude_sec}
Our main conclusions are summarized as followed:

\begin{itemize}
\item{XTE J1550--564 was found to change between spectrally hard and
soft states on time scales of days to weeks. These transition took
place at a more or less constant 2--60 keV count rate level, and were
found at count rate levels that differed by up to a factor 1000.}
\item{As the spectral hardness increased, both the spectral and
variability properties changed from HS via IS to LS. We regard the VHS
as an instance of the IS.}
\item{At least two physical parameters seem to be necessary to account
for the behavior of XTE J1550--564. These parameters vary to a large
extent independently from each other. One of these parameters is
probably the mass accretion rate (through the disk). The other
parameter seems to determine the state of the source and may for
instance be the (relative) size of a Comptonizing region.}

\item{The inner disc radius, as inferred from variability porperties,
increases by a factor of 3--4 as the source moves from the HS to the
LS.}
\item{The properties of the QPOs (frequency, coherence and harmonic
content) as well as the shape and strength of the broad band noise
(weak power law or strong band limited) are well correlated with
spectral hardness.}
\item{The frequencies of the low and high frequency QPOs correlated
well with each other, but in a way that is inconsistent with empirical
relations found for the low and high frequency QPOs in neutron star
systems.}
\end{itemize}

\acknowledgments This research has made use of data obtained through
the High Energy Astrophysics Science Archive Research Center Online
Service, provided by the NASA/Goddard Space Flight Center. This work
was supported by NWO Spinoza grant 08-0 to E.P.J. van den Heuvel, by
the Netherlands Organisation fo Scientific Research (NWO) under
contract number 614-51-002, and by the Netherlands Research-school for
Astronomy (NOVA). RW was supported by NASA through the Chandra
Postdoctoral Fellowship grant number PF9-10010 awarded by the Chandra
X-ray Center, which is operated by the Smitsonian Astrophysical
Observatory for NASA under contract NAS8-39073. MM is a fellow of the
Consejo Nacional de Investigaciones Cient\'{\i}ficas y T\'ecnicas de
la Rep\'ublica Argentina. JH would like to thank Peter Jonker for many
usefull discussions.

\end{document}